# UNIVERSIDADE FEDERAL DO PARANÁ

# Pesquisa sobre a comunicação direta entre usuários e a Atenção Primária à Saúde

## Entendimento de Campanhas no Contexto da Atenção Primária à Saúde: Um Processo de Design Socialmente Consciente

Coordenador: André Grégio

Equipe responsável pelo produto:

- Deógenes Silva Junior
- Jonas Lopes Guerra
- Krissia Menezes
- Marisa Sel Franco
- Roberto Pereira

Curitiba, dezembro de 2024.

# Sumário

# Resumo

*Este relatório apresenta os resultados de uma análise exploratória do contexto de atuação dos Agentes Comunitários de Saúde e Agentes de Combate às Endemias na Atenção Primária à Saúde (APS), especialmente no que tange às Campanhas de Saúde. Para compreender o contexto, foi adotado o referencial de Design Socialmente Consciente, que emprega artefatos e técnicas para examinar o domínio de problemas de maneira abrangente e sociotécnica. Métodos como o Diagrama de Partes Interessadas, Quadro de Avaliação e Framework Semiótico foram utilizados para identificar stakeholders, antecipar desafios e levantar requisitos sociais e técnicos para a solução. Foram aplicadas as técnicas de Personas e Cenários para ilustrar os impactos de uma solução no contexto das campanhas de saúde em diferentes partes interessadas e seus contextos de vida. Este relatório apresenta o método de análise, sua aplicação e resultados, discutindo os achados do estudo para informar a construção de protótipos de média-fidelidade de uma solução para gerenciamento de campanhas de saúde da APS.*

**Palavras-chave:** Campanhas de Saúde, Atenção Primária à Saúde, Design Socialmente Consciente.

# 1. Contexto

O Programa Nacional de Agentes Comunitários de Saúde foi criado em 1991 com o propósito inicial de contribuir para a queda das mortalidades infantil e materna e, em 1994, foi incorporado ao Programa de Saúde da Família (Brasil, 2016). O modelo deste programa é centrado no paradigma da vigilância à saúde e não só visa integrar as políticas públicas com a ação programática em saúde, mas também expandir o lócus de intervenção em saúde porque incorpora o atendimento domiciliar aos pacientes e espaços comunitários (Nunes et al., 2002).

Ainda segundo Nunes et al. (2002), os agentes de saúde, enquanto residentes nas próprias comunidades em que atuam, são os principais interlocutores entre a realidade de saúde de um microterritório e o Sistema Único de Saúde (SUS). Os agentes de saúde possuem, portanto, um papel-chave na tradução do conhecimento técnico para a população e das necessidades da população para os demais profissionais de saúde.

Muitos problemas de saúde são decorrentes de estilos de vida, de comportamentos individuais ou coletivos, em que ações de comunicação de saúde, como o combate ao tabagismo, ou da promoção de atividades físicas, são relevantes para promover saúde e bem-estar amplos para a população (Costa, 2007). Uma das principais ações de comunicação visando a qualidade de vida da população é a campanha de saúde. As campanhas são uma estratégia que orientam para o cuidado com a própria saúde e são aplicadas desde o início do Século XX — por exemplo, nas campanhas contra a febre amarela, em 1903, e contra a peste bubônica, em 1904 (Costa e Carmeiro-Leão, 2021).

Os agentes de saúde possuem um papel fundamental na execução das campanhas da saúde, levando a informação até os domicílios e articulando este conhecimento com os saberes e realidades locais a fim de promover práticas saudáveis e qualidade de vida. Os dados relacionados às iniciativas e execução das campanhas são registrados em sistemas de informação em saúde, como o e-SUS Território.

O e-SUS Território é disponibilizado como aplicativo para dispositivos móveis pelo Departamento de Saúde da Família do Ministério da Saúde como Estratégia e-SUS Atenção Primária à Saúde (e-SUS APS) para registro das ações realizadas por profissionais de saúde em locais de difícil manejo de computadores ou notebooks (Brasil, 2022). O aplicativo tem como principais usuários os Agentes Comunitários de Saúde (ACS), Agentes de Combate às Endemias (ACE) e os Agentes de Ação Social (AAS), dada a capilaridade das ações desses agentes no território. Seu desenvolvimento é integrado ao Sistema e-SUS APS com Prontuário Eletrônico do Cidadão que, entre outras funcionalidades, também possibilita o registro e atualização de terrenos, famílias e cidadãos, ampliando a capacidade de troca de informações entre a equipe.

Dada essa diversidade de agentes envolvidos nas Campanhas da APS, existe a necessidade de que sistemas de informação, como o e-SUS Território, sejam investigados com relação ao apoio que podem fornecer à administração da

APS e aos agentes de saúde no gerenciamento de campanhas. Como parte do Projeto de Pesquisa "*Comunicação direta entre usuários e a Atenção Primária à Saúde (APS)*", há uma demanda pela investigação de uma solução no contexto das campanhas de saúde, apoiando as partes interessadas deste contexto a criar, gerenciar e monitorar campanhas no território brasileiro.

As campanhas são cruciais por constituírem a principal estratégia de comunicação em saúde. Campanhas de saúde possuem características próprias, como o fato de envolverem a atuação de diferentes profissionais da APS, além de precisarem atingir partes interessadas de diversas regiões do Brasil, com necessidades e desafios específicos de suas únicas configurações geográficas, socioeconômicas e culturais. Por estas características, o problema de gerenciamento de campanhas do SUS é complexo. É necessária uma abordagem multidisciplinar para entender, formalizar, documentar e ser capaz de lidar com esse cenário complexo, assim como prospectar uma solução computacional para apoiar o trabalho dos gerentes e agentes de saúde. Por isso, esta pesquisa combina métodos de investigação da literatura com métodos de design de sistemas computacionais informados por uma perspectiva de design socialmente consciente (Baranauskas et al., 2024). Essa abordagem multidisciplinar permite considerar de forma integrada aspectos técnicos do problema/solução, bem como necessidades e desafios enfrentados pela ampla diversidade de partes interessadas envolvidas na Atenção Primária à Saúde.

Como Produto do Projeto de Pesquisa "*Comunicação direta entre usuários e a Atenção Primária à Saúde (APS)*", este relatório técnico apresenta um diagnóstico elaborado a partir de sete atividades principais:

1. Investigação sobre o contexto da APS a partir de uma Revisão bibliográfica exploratória e de uma Pesquisa Netnográfica;
2. Identificação dos Stakeholders (partes interessadas) envolvidos no Diagrama de Partes interessadas;
3. Identificação de desafios, dificuldades e barreiras do partes interessadas no contexto de Campanhas de Saúde por meio do artefato Quadro de Avaliação;
4. Síntese dos desafios, declaração do problema e prospecção de uma solução de apoio e gerenciamento de campanhas de saúde a partir do artefato Quadro de Prospecção de Valor;
5. Representação de personas e cenários de interação;
6. Definição de Requisitos Prospectivos da Solução por meio do Framework Semiótico;
7. Elaboração de Protótipos de Média Fidelidade ilustrando a solução de gerenciamento de campanhas.

As sete atividades contribuem para o desenvolvimento da Meta 5 "*Propor uma solução computacional para a interação direta com os usuários em piloto realizado em municípios indicados pelo Ministério da Saúde (MS) considerando as equipes de Atenção Primária à Saúde*", especialmente da Meta 5.1 - "*Conduzir* *atividades para a identificação das necessidades e desafios de interação dos usuários*". Como principais resultados, foram identificadas 60 (sessenta) partes interessadas diversas, assim como desafios e problemas que essas partes possuem no contexto das Campanhas de Saúde na APS. Um total de 35 (trinta e cinco) histórias de usuário organizadas em diferentes níveis de formalidade foram especificadas, detalhando requisitos para uma solução para o gerenciamento e apoio na execução de Campanhas. Por fim, sete protótipos de média fidelidade foram desenvolvidos, materializando a proposta de solução para o gerenciamento de campanhas de saúde na perspectiva de gerentes da APS.

As próximas seções deste relatório estão organizadas da seguinte maneira: a Seção 2 apresenta os conceitos de APS, os agentes de saúde e o contexto das Campanhas de Saúde. A Seção 3 apresenta os materiais e métodos utilizados como estratégia de investigação desta pesquisa, sendo primariamente o Design Socialmente Consciente e demais métodos de coleta de dados. A Seção 4 apresenta os resultados obtidos a partir do entendimento do contexto e da modelagem e prototipação da solução. Por fim, a Seção 5 apresenta as considerações finais.

## 2. Fundamentação

A APS universal, integral e equitativa é o modelo de atenção mundialmente qualificado para alcançar um nível de saúde que permita à população levar uma vida social e economicamente ativa (Facchini et al., 2018). No Brasil, a estruturação de serviços de APS é um importante diferencial para a consolidação dos princípios e diretrizes do SUS e para a obtenção de melhores resultados em prol de um cuidado mais equânime, abrangente e universal (Castanheira et al., 2024). Estudos sobre a APS evidenciam seu impacto na saúde e no desenvolvimento da população nos países que a adotaram como base para seus sistemas de saúde: melhores indicadores de saúde, maior eficiência no fluxo dos usuários dentro do sistema, tratamento mais efetivo de condições crônicas, maior eficiência do cuidado, maior utilização de práticas preventivas, maior satisfação dos usuários, e diminuição das iniquidades sobre o acesso aos serviços e o estado geral de saúde (Oliveira e Pereira, 2013).

A APS tem sua importância reconhecida como uma das estratégias mais efetivas na redução de mortes e internações por vários agravos e doenças, principalmente doenças crônicas não transmissíveis (Barros et al., 2024), priorizando ações de promoção, proteção e recuperação de saúde, de forma integral e continuada (Oliveira e Pereira, 2013). Dentro da APS, as campanhas de saúde são centrais como meio de comunicação e divulgação de informações, assim como a realização de ações concretas (ex., vacinação, exames) que visam apoiar a vigilância e promoção de qualidade de vida para a população. Agentes de Saúde, como ACS e ACE, possuem papel direto nas campanhas, sendo mais que apenas transmissores de informação. Esses agentes possuem um papel importante na articulação da campanha: por um lado, auxiliando as equipes de saúde a identificarem e encaminharem famílias com situações críticas em relação ao objetivo de uma campanha; por outro lado, possibilitando que a população conheça, crie sentido e promova mudanças concretas de qualidade de vida relacionadas à campanha. Deste modo, esta seção fundamenta os principais conceitos desta pesquisa: a Subseção 2.1 fundamenta o papel destes agentes no contexto da APS, inclusive nas ações de comunicação em saúde, enquanto a Subseção 2.2 fundamenta o conceito de campanhas de saúde.

## 2.1 Agentes Comunitários de Saúde e Agentes de Combate às Endemias

A atuação multiprofissional da APS foi um diferencial na política de saúde (Facchini et al., 2018), conjugando diferentes profissionais, como médicos, enfermeiros, profissionais de odontologia, técnicos e ACS, em uma atuação conjunta. O fortalecimento dessa atuação em equipe e do papel dos agentes de saúde é essencial para garantir a integralidade e a coordenação das ações de comunicação e promoção da saúde, prevenção de doenças e cuidado de problemas e agravos clínicos, de cada usuário do serviço (Facchini et al., 2018).

Contextos e situações de vulnerabilidades e riscos socioambientais, que geram adoecimentos e agravos à saúde, demandam ações articuladas e integradas de vigilância e atenção à saúde (de Araújo e de Sousa, 2021). A integração da APS e Vigilância Sanitária demanda, em especial, a identificação de processos socioambientais geradores de condições que favorecem a exposição a fatores de risco e o adoecimento da população (de Araújo e de Sousa, 2021). Muitas dessas condições se tornam objetivos de campanha, como o combate a arboviroses. Na relação entre a APS e a vigilância sanitária, o ACS e o ACE desempenham papéis centrais no contato e comunicação com a população.

O ACS tem um papel fundamental no cuidado em saúde no contexto da APS, estando centralmente envolvido no conjunto de ações desenvolvido no âmbito da Estratégia Saúde da Família (ESF), especialmente no que tange à promoção e à proteção da saúde, a prevenção de agravos, o auxílio em ações de diagnóstico, tratamento, reabilitação e manutenção da saúde. Esses profissionais participam de ações próximas de comunidades locais. Visitas às famílias, panfletagem de campanhas de saúde em locais estratégicos, vacinação, aferição de pressão, acompanhamento de pacientes acamados ou com mobilidade reduzida são exemplos de atividades realizadas por ACS. Os ACS representam o mais forte ponto da conexão entre a APS e a vigilância em saúde, considerando seu vínculo com as famílias no território (Schenkman et al., 2023). No contato com moradores, ACS identificam situações de saúde muitas vezes não manifestadas como demandas, que só chegam ao sistema de saúde pela atuação desses profissionais (Morosini e Fonseca, 2018).

A visita domiciliar, que consiste no acompanhamento das condições de saúde das famílias de sua microárea e na busca ativa de situações específicas, é uma das atividades preponderantes e principal expressão da presença do ACS no território (Morosini e Fonseca, 2018). As visitas domiciliares também são um dos principais caminhos para o alcance e divulgação das campanhas de saúde, já que a mensagem pode ser comunicada diretamente no local onde as pessoas vivem.

No contexto da Atenção Primária, a Vigilância em Saúde também é importante para entender fatores de risco à saúde e processos de adoecimento da população, por exemplo relacionado às arboviroses como Dengue, Chikungunya e Zika. Os ACE realizam ações de vigilância em saúde, integrando os processos de trabalho, planejamento, programação, monitoramento e avaliação de forma integrada e complementar, fortalecendo ações de vigilância e promoção da saúde na APS (SBIBAE, 2019).

O ACE é responsável pelo monitoramento e controle de doenças transmitidas por vetores, como a dengue, Zika, chikungunya, malária, bem como ações que incluem a promoção de ambientes saudáveis, vigilância e identificação de riscos ambientais que possam impactar negativamente a saúde (Rodrigues, 2024). A atuação dos ACS encontra sinergias com a atuação dos ACE, em que essa atuação comum entre os dois tipos de profissionais é fundamental para a territorialização, o diagnóstico das condições de saúde e de vida da população na APS (de Araújo e de Souza, 2021).

Os ACE têm, dentre suas atribuições, a realização de visitas domiciliares, com o foco na vigilância em saúde. Muitas campanhas de enfrentamento às arboviroses são veiculadas anualmente, nas quais ACE possuem papel central em visitas domiciliares na articulação de conhecimentos sobre vetores e sintomas de doenças, assim como realizando e orientando ações práticas importantes no trabalho de prevenção. Somente nos dois últimos anos (2013-2014)[1], foram veiculadas cinco campanhas de saúde em relação às arboviroses combatidas especialmente pelos ACE. As visitas são uma oportunidade para a detecção de riscos e vulnerabilidades que possam gerar processos de adoecimento ou agravos à saúde no território e também para a construção de vínculo familiar e comunitário (de Araújo e de Souza, 2021).

## 2.2 Campanhas de Saúde

A importante relação entre comunicação e saúde é anterior à própria criação do SUS. Comunicação e saúde são dois campos transdisciplinares que inicialmente trabalharam as propagandas e campanhas sanitárias visando à mudança de estilos de vida (Brito et al., 2020).

De acordo com Brito et al. (2020), a comunicação em saúde é um processo transacional e fundamental às atividades de promoção da saúde, no qual em um modelo de comunicação em saúde efetivo, presume-se que há um emissor (geralmente um profissional de saúde) que envia uma mensagem a um receptor (usuários dos serviços de saúde), há o entendimento da mensagem e o retorno do receptor, em uma via bidirecional.

Campanhas de saúde podem, por exemplo, universalizar os cuidados de pós-parto, a imunização de crianças, gestantes e idosos, o rastreamento de câncer de colo de útero e de mama, o cuidado dos pés de pessoas com diabetes, a avaliação de risco cardiovascular em adultos e idosos, a promoção de alimentação saudável e de atividade física (Facchini et al., 2018).

---

[1] Último acesso em: 17/12/2024. Disponível em: https://www.gov.br/saude/pt-br/campanhas-da-saude

A Política Nacional de Atenção Básica (Brasil, 2017) define atribuições compartilhadas entre ACS e ACE no contexto da comunicação em saúde, como orientar a comunidade sobre sintomas, riscos e agentes transmissores de doenças e medidas de prevenção individual e coletiva. Também existem as atribuições de informar e mobilizar a comunidade para desenvolver medidas simples de manejo ambiental e outras formas de intervenção no ambiente para o controle de vetores.

As campanhas de saúde envolvem planejamento para sua execução, pois abrangem um calendário determinado, orçamento, metas e indicadores, assim como insumos associados (ex: vacinas). Em relação à vacinação, por exemplo, o Governo Federal publicou um manual de microplanejamento para as atividades de vacinação, definindo quatro etapas de planejamento para uma vacinação de alta qualidade (BRASIL, 2023).

Ainda em relação às campanhas de vacinação, em 2023 o Ministério da Saúde mudou a estratégia nacional de imunização, adicionando um calendário diferente para atender às particularidades climáticas da região norte. Ao mesmo tempo que flexibilizar as campanhas pode auxiliar no maior impacto das campanhas, também aumenta a complexidade para gestores e executores realizarem o gerenciamento e aplicação dessas campanhas a nível nacional.

Além dessa mudança da estratégia nacional de imunização, localmente nos municípios os ACS e demais profissionais das Equipes de Saúde da Família têm papel fundamental nas campanhas. Entre as estratégias que podem ser adotadas, tem-se a realização do Dia D de vacinação, busca ativa de não vacinados, vacinação nas escolas, vacinação para além das unidades de saúde, checagem da caderneta de vacinação e intensificação da vacinação em áreas indígenas (BRASIL, 2024). Deste modo, a campanha não significa apenas a distribuição de material informativo (ex., panfletos e cartazes), mas envolve um conjunto de ações e estratégias amplas de promoção à saúde, que por sua vez também adicionam maior esforço e complexidade na articulação entre as equipes na execução das atividades.

As campanhas de saúde ainda se tornam mais relevantes em um contexto social onde as *fake news* e a desinformação afligem constantemente a população com informações anti-vacinas (de Barcelos et al., 2021; Vasconcelos-Silva e Castiel, 2020; Tokojima Machado et al., 2020), a promoção de comportamentos maléficos para a saúde (Irfan et al., 2024; Elkhazeen et al., 2023) e a divulgação de tratamentos de saúde não efetivos (Domaradzki et al., 2024; Leonard e Philippe, 2021). O desafio da desinformação reforça ainda mais a importância das campanhas, que precisam ser planejadas com antecedência, prevendo o manejo de recursos e as estratégias de divulgação e comunicação a serem realizadas pela equipe da APS. O manual de microplanejamento para atividades de vacinação, por exemplo, determina como ponto importante a ser considerado a preparação dos materiais das campanhas de vacinação pelo menos três meses antes das ações específicas, para permitir imprimi-los e distribuí-los em tempo hábil (BRASIL, 2023).

Considerando as desigualdades e a extensão territorial brasileira, as tecnologias, o planejamento e execução das campanhas também pode se beneficiar do apoio de tecnologias. As tecnologias são de grande valia para a resolutividade e ampliação do acesso à saúde com qualidade à população brasileira (Bousquat et al., 2017), produzindo estimativas essenciais, como o da cobertura de ações e programas de saúde (Facchini et al., 2018). Durante a pandemia, agentes de saúde utilizavam canais de comunicação instantâneos para fornecer orientações e práticas de saúde preventivas da COVID-19 (Schenkman et al., 2023).

A expansão da digitalização no contexto da saúde é relevante, desde que realizada de forma consciente e inserida nas realidades socioeconômicas e culturais locais. No projeto de Pesquisa "Comunicação direta entre usuários e a Atenção Primária à Saúde (APS)", foi definida uma demanda de investigação sobre tecnologias que apoiem o gerenciamento de campanhas de saúde. Visando realizar um processo de design socialmente consciente, foi planejada a estratégia de investigação descrita na seção a seguir, com o propósito de compreender o contexto das campanhas e fundamentar o projeto de uma solução para apoiar a criação e gerenciamento de campanhas de saúde.

## 3. Método

Como estratégia de investigação para a condução deste estudo foram realizadas: 1) uma revisão bibliográfica exploratória (Marconi e Lakatos, 2004) e 2) uma netnografia (Hussein et al., 2016) de forma a coletar dados para informar o entendimento sobre o contexto da APS, assim como a atuação dos agentes de saúde; 3) foram utilizados artefatos do Design Socialmente Consciente (Baranauskas et al., 2024) para entender o problema de forma abrangente, sistemática e teoricamente informada, considerando tanto questões sociais quanto formais e técnicas relevantes para a APS e para a atuação de ACS e ACE; 4) foram utilizadas as técnicas de Personas e Cenários para apoiar a representação do entendimento do problema; e 5) foram construídos protótipos de média fidelidade para fundamentar a construção de uma solução para o gerenciamento de campanhas. A Figura 1 a seguir representa uma visão geral da estratégia de investigação desta pesquisa.

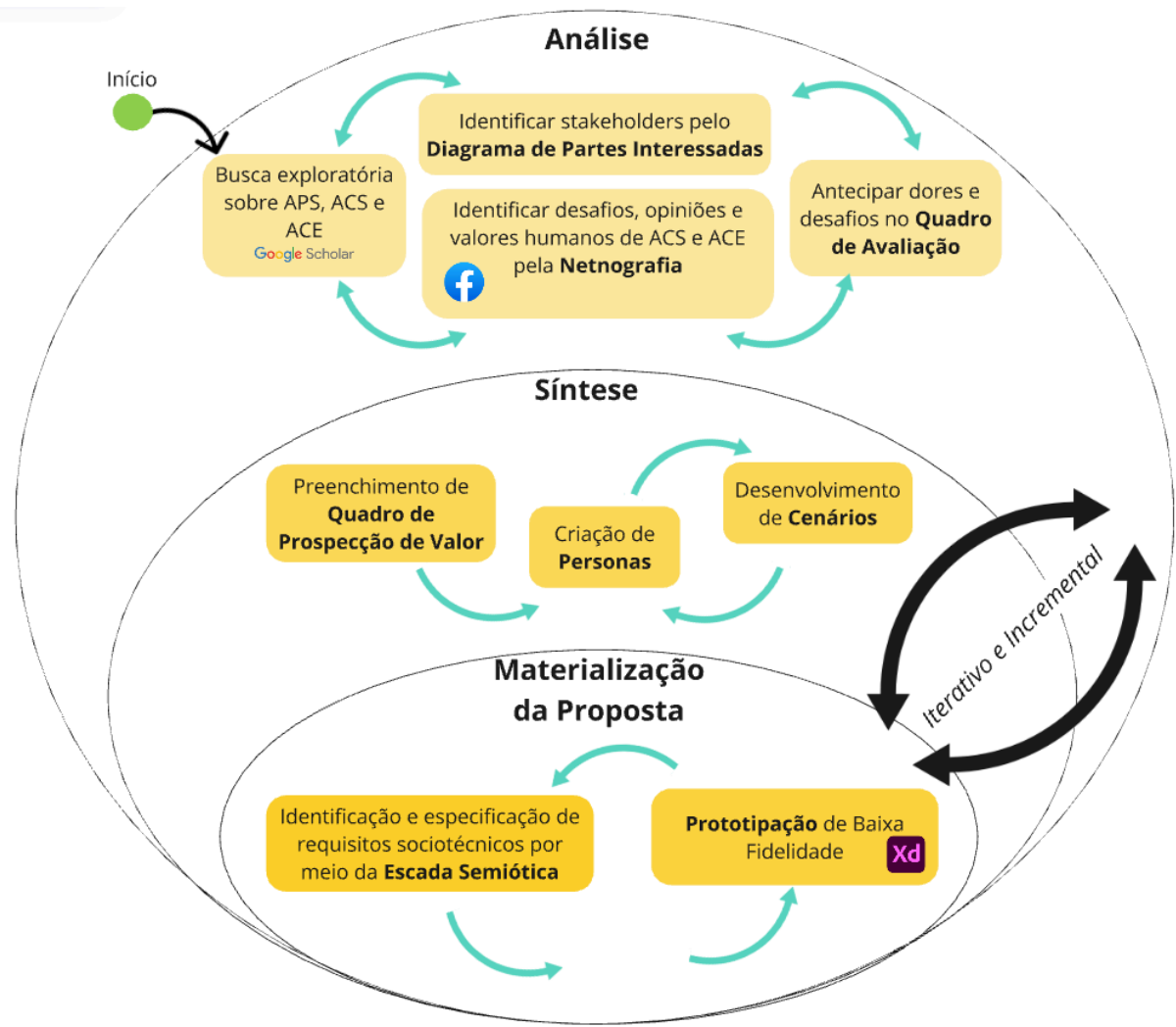

**Figura 1.** Etapas e atividades que compõem a estratégia de investigação desta pesquisa. As etapas do Design Socialmente Consciente envolvem a Análise, Síntese e Materialização da Proposta. Em cada etapa, atividades alinhadas ao objetivo de cada etapa foram realizadas. O processo foi realizado de forma iterativa e incremental, de modo que o conhecimento adquirido ao longo da pesquisa foi utilizado para refinar os resultados de etapas anteriores.

O Design Socialmente Consciente (Baranauskas et al., 2024) é um *framework* de design que traz preocupações com aspectos socioculturais tanto no entendimento de um contexto quanto de uma tecnologia que vai operar neste contexto. Esse *framework* traz 3 etapas macro: 1. Análise, 2. Síntese e 3. Materialização da Proposta (ou da solução), oferecendo diversos artefatos, métodos e técnicas para apoiar os projetistas a conduzir um processo de design consciente de aspectos informais e sociais de um domínio. Deste modo, é uma abordagem adequada para a análise de contextos socioculturalmente complexos, como o da saúde brasileira, que envolvem partes interessadas com diferentes condições econômicas, culturais e sociais. Na sequência, apresentamos as principais atividades conduzidas em cada etapa.

*1. Análise.* O objetivo foi entender, no contexto das campanhas de saúde da APS, quem eram as partes interessadas afetadas, seus principais desafios, barreiras e problemas de forma abrangente. Este é um objetivo relevante na pesquisa, visto a dimensão geográfica extensa do Brasil e diferenças socioculturais encontradas. A situação complexa brasileira indica previamente que a situação de saúde e as estratégias de campanhas de saúde não são iguais em todo o país, assim como não serão iguais as pessoas e suas necessidades. Para identificar os stakeholders em abrangência, foi utilizado o artefato **Diagrama de Partes Interessadas**, que apresenta cinco níveis de influência em que partes interessadas influenciam ou são influenciadas pelo problema ou sua solução. O **Quadro de Avaliação**, por sua vez, estende o Diagrama de Partes Interessadas para entender os desafios, questões e problemas dessas partes interessadas e antecipar ideias para solucionar os desafios mapeados.

Como forma de entender o contexto, foram coletados dados em **buscas exploratórias** utilizando o *Google Scholar*, uma ferramenta que permite identificar artigos relevantes, incluindo bases digitais da saúde, ciências sociais, entre outros. Os artigos foram selecionados para leitura caso contribuíssem com o entendimento sobre o contexto das campanhas de saúde da APS e sobre a atuação de ACS e ACE neste contexto. Além de informar o preenchimento dos artefatos e prototipação desta pesquisa, os artigos selecionados compõem a Seção 2 de Fundamentação Teórica.

Com o mesmo objetivo, foi realizada uma **Netnografia** na rede social Facebook. A Netnografia é uma forma de etnografia digital, investigando em postagens, grupos e comunidades digitais sobre as práticas, opiniões e comportamentos que as partes interessadas compartilham publicamente *online*. A partir de uma Netnografia, pesquisadores podem se aproximar da linguagem, dores e desafios que essas partes interessadas compartilham e expressam espontaneamente em seus ambientes digitais. Um relatório da Netnografia realizada está disponível no Apêndice 2 deste relatório.

*2. Síntese.* O objetivo foi sintetizar o entendimento produzido e retomar as necessidades das principais partes interessadas quando se trata das campanhas de saúde da APS. Deste modo, declara-se explicitamente os desafios, quem é afetado pelo problema e em qual contexto específico ele ocorre.

Nesta etapa, as informações da etapa anterior, coletadas em abrangência, foram sintetizadas visando se aproximar do contexto de uma provável solução de gerenciamento e criação de campanhas de saúde a partir do Quadro de Prospecção de Valor (Ferrari et al., 2020). Este artefato é relevante em contextos em que pode-se perder de vista as partes interessadas principais de um projeto. Há um risco em projetos de design que o stakeholder principal seja ignorado, quando comparado aos valores, desejos e expectativas de stakeholders de gestão, administração e envolvidos no desenvolvimento das tecnologias. Deste modo, o Quadro de Prospecção de Valor foi utilizado para colocar o stakeholder principal no centro do processo de design, entendendo qual é sua principal necessidade e como a solução sendo desenvolvida pode inovar e surpreender essa parte interessada.

A partir desse artefato, Personas (Barbosa et al., 2021) foram desenvolvidas para materializar as partes interessadas com nomes, lugar de moradia no Brasil, seus desafios e necessidades. Com isso, traz-se à tona os diversos desafios socioculturais presentes em diferentes perfis demográficos do Brasil. A técnica de *persona* foi utilizada combinada à técnica de cenários (Rosson e Carroll, 2012), em que a *persona* foi utilizada como personagem principal de cenários que descreviam as necessidades da parte interessada, o contexto de saúde, e como o e-SUS Território se relacionava com esse contexto.

*3. Materialização da Proposta.* O objetivo foi identificar e especificar requisitos, assim como prototipar a solução sendo proposta, referente à extensão do e-SUS Território para considerar campanhas de saúde e seu gerenciamento. Essa especificação e prototipação foi realizada à luz dos conhecimentos, empatia e consciência sociocultural adquiridas nas etapas anteriores.

Nesta etapa, utilizou-se do Framework Semiótico (Stamper, 1993) para apoiar a identificação de requisitos sociotécnicos de uma solução de gerenciamento e criação de para campanhas de saúde. O Framework Semiótico apresenta seis níveis que são utilizados para organizar qualquer sistema de informação. Dentre eles, os três níveis superiores são relacionados ao sistema de informação humano, envolvendo aspectos sociais, de intenções e significados; os três níveis inferiores são relacionados ao sistema técnico, tratando de aspectos de forma, estrutura, frequência e infraestrutura técnica.

A partir do entendimento do contexto e do delineamento dos requisitos para a solução prospectiva, foi realizada uma prototipação de média fidelidade ilustrando visualmente a interface da solução e seus fluxos de interação. O Adobe XD[2] foi utilizado como ferramenta para apoiar a prototipação.

## 4. Resultados

Os resultados estão organizados a partir de cada artefato, sendo que estes foram preenchidos a partir das evidências e indícios adquiridos com o uso das técnicas de coleta de dados (netnografia e leitura documental).

### 4.1 Diagrama de Partes Interessadas

Utilizando como base o Diagrama de Partes Interessadas da Semiótica Organizacional (Stamper et. al, 2000), a Figura 2 mostra as partes interessadas identificadas distribuídas nas diferentes camadas do diagrama, representando a sua proximidade com o problema em análise. Criado com base na Teoria de Stakeholders (Donaldson e Preston, 1995), o diagrama chama a atenção para a diversidade de partes interessadas que influenciam ou são influenciadas pelo problema central em análise: comunicação de campanhas do Ministério da Saúde aos ACS e aos ACE. A camada de operação representa o objeto de análise em questão. A camada de contribuição representa as partes interessadas que influenciam ou são influenciadas diretamente pelo problema ou solução. A camada de fonte representa as partes que são fonte de informação sobre o contexto. A camada de mercado, por sua vez, representa parceiros e concorrentes. Por fim, a camada de comunidade é que possui o menor nível de influência, representando legisladores e espectadores do contexto.

---

[2] Disponível em: <https://adobexdplatform.com/>. Último acesso em 11/11/2024.

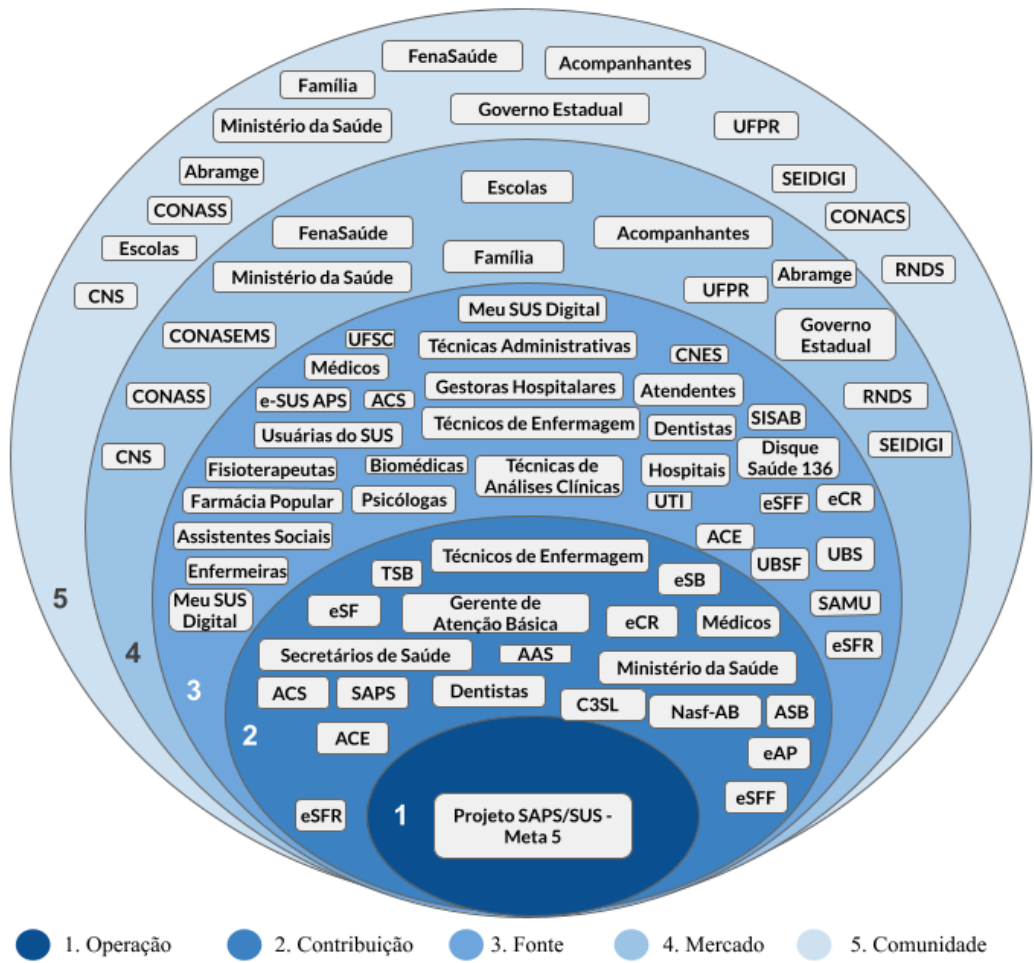

**Figura 2.** Diagrama de Partes Interessadas no Contexto de Campanhas na Atenção Primária à Saúde.

A Figura 2 representa as mais de 60 partes interessadas no contexto de Campanhas na APS. A abrangência de partes interessadas mostra o desafio desta pesquisa em investigar um sistema para a saúde pública, pois coexistem partes interessadas a nível federal (Ministério da Saúde, CNS, RDS, SISAB), estadual (Governos e Escolas Estaduais), municipal (equipes da eSF, eSFF, eSFR) e regional (ACS e ACE que atuam em territórios locais, por exemplo).

Identificou-se a diversidade de condições geográficas, culturais e socioeconômicas que pessoas usuárias do SUS possuem no Brasil, um país de dimensões continentais, com uma ampla variedade de recursos, condições de acesso e infraestrutura. No norte, por exemplo, há usuários do SUS que vivem em regiões ribeirinhas, com condições de acesso e comunicação de saúde especiais para este tipo de parte interessada. Há usuários do SUS de zonas rurais, quilombolas e diferentes outros tipos de configurações geográficas e socioeconômicas. Do mesmo modo, co-existem equipes de saúde para condições geográficas (ex., eSFF para famílias fluviais) e socioeconômicas específicas (eCR para pessoas em situação de rua). Este é um indicativo que uma campanha de saúde também deve se apropriar dessas especificidades em seu planejamento e execução, pois a abordagem de comunicação em saúde pode possuir demandas diferentes dependendo do contexto a qual está sendo executada.

      Também é possível observar uma diversidade de atores na APS que são responsáveis pela estratégia de comunicação em saúde, desde partes interessadas na gerência da APS, os membros da Equipe APS e da Equipe de Saúde da Família, bem como agentes de saúde que atuam no território, interagindo diretamente com usuários, famílias, escolas e outros espaços sociais onde as pessoas convivem. As diferentes partes interessadas identificadas mostram que, no Brasil, a saúde se faz em rede, de forma relacional entre diferentes profissionais (ACE, ACS, Técnicos, Médicos, Dentistas, etc.), que por sua vez formam diferentes equipes (eSF, eAP, eSB, eSFF, eSFR, eCR). Deste modo, identifica-se a importância de se considerar partes interessadas indiretas (aquelas que não usarão uma solução diretamente), pois entende-se possíveis influências que um conjunto de partes interessadas pode introduzir no uso de uma solução. Visto que a saúde ocorre em rede e em colaboração entre diferentes partes interessadas, uma solução no contexto das campanhas pode maximizar as sinergias e parcerias entre as partes interessadas e não negligenciar as relações profissionais e sociais que existem neste contexto.

      Com a análise, verifica-se que a APS não fica confinada às unidades de saúde (UBS e UBSF), mas tem uma inserção na vida e nos espaços sociais das pessoas, nos quais diversas outras partes interessadas são impactadas (Famílias, Hospitais, Farmácias e Escolas). O contexto de investigação não deve então ficar restrito aos contextos hospitalares, apenas, mas adentrar o mundo social em que as pessoas vivem e se relacionam em contextos de saúde.

      Diversos conselhos e outros órgãos legisladores foram identificados (CNS, CONACS, CONASEMS), que por sua vez definem normas, direitos e deveres e têm papel fundamental na discussão sobre a APS no Brasil. Estas partes interessadas podem definir requisitos importantes para qualquer solução digital no contexto das campanhas, principalmente sobre aspectos formais de um sistema (normas e legislações que devem ser considerados). Partes interessadas relacionadas ao contexto digital em saúde (Meu SUS Digital, Disque Saúde 136, e-SUS APS, SISAB) informam uma rede de atuação digital de saúde, assim como canais e sistemas que devem ser considerados ao menos em termos de interoperabilidade técnica (troca e compartilhamento de informações).

## 4.2 Quadro de Avaliação

O Quadro de Avaliação teve seu preenchimento baseado em evidências de estudos científicos no contexto da Atenção Primária à Saúde (Barros et al., 2024; Batistella, 2013; Bousquat et al., 2017; Brito, 2020; Castanheira et al., 2024; de Araújo et al., 2021; Evangelista et al., 2018; Facchini et al., 2018; Morosini e Fonseca, 2018;

Nuances et al., 2002; Pessoa et al., 2016; Schönholzer et al., 2021) e netnografia em redes sociais em grupos colaborativos online sobre ACS e ACE.

Foi seguida a recomendação da literatura de garantir que a parte interessada mais crítica ou importante de cada camada seja incluída na análise do Quadro de Avaliação (Pereira e Baranauskas, 2015). Ao garantir que partes interessadas de cada camada sejam consideradas, evita-se ignorar informação importante que pode afetar o design do problema ou a sua prospectiva solução (Ferrari et al., 2020). Os problemas, necessidades, desafios identificados, assim como a prospecção de ideias de solução para estes desafios estão mapeados e detalhados no Apêndice 1 e são exemplificados a seguir de acordo com as partes interessadas de cada camada.

**4.2.1 Contribuição**

Foram identificadas cinco principais partes interessadas na camada de Contribuição, sendo elas: 1. Pessoa Usuária do SUS, 2. Profissionais da Atenção Primária à Saúde (de forma geral), 3. Agentes Comunitários de Saúde, 4. Agentes de Combate à Endemias e 5. Gestores Municipais da Atenção Primária à Saúde. A seguir são apresentados exemplos de problemas e questões e ideias de solução para cada uma dessas partes interessadas.

*1. Pessoa Usuária do SUS:* esta parte interessada enfrenta problemas como a falta de acesso a meios de comunicação de massa e a dificuldade de compreensão das instruções de campanhas de saúde, perdendo prazos, deixando de atender a chamados do SUS e/ou de tomar atitudes simples relacionadas a essas iniciativas. Facilitar o processo de comunicação das campanhas de saúde aos usuários do SUS, provendo diferentes recursos para transmissão das mensagens de forma inclusiva, acessível e com linguagem simples são ideias de soluções para estes problemas.

*2. Profissionais da Atenção Primária à Saúde*: a dificuldade para operar os sistemas do SUS, com problemas para manter registros de atendimentos atualizados e corretos, curva de aprendizagem complexa para aprender a usar os sistemas e dificuldades de uso e de acesso por diferentes questões (ex.: qualidade do acesso à Internet, habilidades de uso, etc.) são exemplos de problemas identificados para esta parte interessada. Projetar soluções que funcionem no modo *offline*, que possam transmitir dados apenas quando conectadas em Wi-Fi, e que operem em dispositivos com baixa capacidade de processamento e armazenamento são ideias de soluções para estes problemas.

*3. Agentes Comunitários de Saúde:* ACS enfrentam problemas em criar vínculos com famílias, devido a muitas mudanças domiciliares em suas microáreas e a dificuldade de encontrar famílias no horário de atuação, tendo que atualizar constantemente os dados demográficos de moradores. Funcionalidades que facilitem a criação de vínculos com famílias, de deletar e atualizar um domicílio nos sistemas de informação de saúde utilizados são ideias de soluções para estes problemas.

*4. Agentes de Combate às Endemias:* percorrer longas distâncias sob sol, chuva, frio carregando materiais a pé ou utilizando recursos pessoais (ex.: bicicleta) são exemplos de problemas enfrentados por esta parte interessada. Incluir recursos de deslocamento no planejamento de trabalho do ACE são ideias de soluções para estes problemas.

*5. Gestores Municipais da Atenção Primária à Saúde:* esta parte interessada enfrenta problemas como a dificuldades no planejamento das ações da APS, bem como do gerenciamento e acompanhamento da atuação da equipe. Auxiliar no planejamento local da execução da Campanha, utilizando ou estabelecendo padrões, como o da vacinação de alta qualidade (BRASIL, 2023) e uma solução que permite monitorar indicadores relacionados à execução da Campanha e do trabalho da equipe são ideias de soluções para estes problemas.

### 4.2.2 Fonte

Foram identificadas três principais partes interessadas na camada Fonte, sendo elas: 1. Unidades Básicas de Saúde, 2. Farmácia Popular e 3. Meu SUS Digital. A seguir são apresentados exemplos de problemas e questões e ideias de solução para cada uma dessas partes interessadas.

*1. Unidades Básicas de Saúde:* enfrenta problemas com sistemas não integrados à Rede Nacional de Dados em Saúde (RNDS). *O* (re)design de soluções para adotarem padrões de interoperabilidade brasileira são ideias de soluções para estes problemas.

*2. Farmácia Popular:* Irregularidades nas vendas de medicamentos, controle de estoque deficitário e sucateamento de material permanente são exemplos de problemas enfrentados por essa parte interessada. A definição de indicadores de ações fraudulentas, gerenciamento de recursos no SUS e monitoramento destes indicadores são ideias de soluções para estes problemas.

*3. Meu SUS Digital:* Os desafios técnicos de escalabilidade e interoperabilidade entre as aplicações vinculadas ao Meu SUS são exemplos de problemas enfrentados por essa parte interessada. A adoção de padrões de interoperabilidade em soluções novas ou redesign de soluções existentes a partir destes padrões são ideias de soluções para estes problemas.

### 4.2.3 Mercado

Foram identificadas duas principais partes interessadas na camada Mercado, sendo elas: 1. Secretarias Municipais de Saúde e 2. Prefeituras. A seguir são apresentados exemplos de problemas e questões e ideias de solução para cada uma dessas partes interessadas.

*1. Secretarias Municipais de Saúde:* Avaliação dos serviços prestados via SUS e a operação complexa dos sistemas de informações do SUS são exemplos de problemas enfrentados por essa parte interessada. Uma solução para receber feedback sobre atendimento no SUS, com diferentes modalidades para atender diferentes condições físicas e socioeconômicas é uma ideia de solução para estes problemas.

*2. Prefeituras:* enfrenta problemas em fazer um acompanhamento adequado das pessoas em tratamento pelo SUS. Um Sistema de informação de saúde com acompanhamento fino da jornada e situação de saúde dos cidadãos é uma ideia de solução para estes problemas.

### 4.2.4 Comunidade

Foram identificadas cinco principais partes interessadas na camada Comunidade, sendo elas: 1. Família, 2. Acompanhantes, 3. Ministério da Saúde, 4. Estados e 5. Municípios A seguir são apresentados exemplos de problemas e questões e ideias de solução para cada uma dessas partes interessadas

*1. Família:* enfrenta problemas como a falta de privacidade e de individualidade no sistema de saúde. Viabilizar que um mesmo dispositivo tecnológico seja utilizado para o acompanhamento de mais de uma pessoa, permitir o acesso e gerenciamento a aplicativos como MeuSUS por pais ou responsáveis de dependentes, garantir a privacidade e o tratamento adequado de dados pessoais e dados sensíveis são ideias de soluções para estes problemas.

*2. Acompanhantes:* Sofrimento psíquico sobre as dores, indefinição ou tratamento do paciente acompanhado e ser tratado de forma não humanizada são exemplos de problemas enfrentados por essa parte interessada. A aplicação e monitoramento de protocolos de atendimento humanizados que reduz, pela empatia e acolhida, o sofrimento de pacientes e acompanhantes é uma ideia de solução para esses problemas.

*3. Ministério da Saúde:* Os desafios para a articulação entre os programas de saúde e a sociedade, a concentração e falta de profissionais de saúde pelo território nacional, as dificuldades em atender um país de dimensões continentais e as grandes diferenças regionais que requerem atenção são exemplos de problemas enfrentados por essa parte interessada. Projetar uma solução que possa se comunicar com diferentes sistemas nos níveis federal, estadual e municipal, respeitando padrões de interoperabilidade, software livre e gratuito são ideias de solução para estes problemas.

*4. Estados:* enfrenta problemas como desafios com o aprisionamento tecnológico, na manutenção da soberania tecnocientífica e necessidade de soluções que garantam a autonomia e a segurança no contato com as pessoas e na interação com seus dados. Utilizar soluções baseadas em software livre e infraestrutura baseada em modelos, protocolos e tecnologias abertas, gratuitas e livres são ideias de solução para estes problemas.

*5. Municípios:* Desafios em acompanhar avanços tecnológicos são exemplos de problemas enfrentados por essa parte interessada. Utilizar soluções baseadas em software livre e infraestrutura baseada em modelos, protocolos e tecnologias abertas, gratuitas e livres são ideias de solução para estes problemas.

Como principais resultados, identificou-se para usuários do SUS principalmente desafios de acesso a serviços de saúde e de entendimento das campanhas, por sua condição socioeconômica (ex., não possuir dispositivo tecnológico ou ter acesso limitado à internet) e localização geográfica. Principalmente para as pessoas que não possuem acesso a canais de informação, a mensagem de comunicação em saúde pode ficar ainda mais enfraquecida, justamente para pessoas que podem viver em condições de riscos e vulnerabilidades em saúde. No caso da febre amarela no Brasil, por exemplo, o ciclo da doença atualmente é silvestre (zonas rurais ou de floresta), sendo que os últimos casos de febre amarela urbana foram registrados em 1942[3]. Barreiras como falta ou limitação de internet, não acesso a dispositivos tecnológicos e longa distância de agentes de saúde podem exacerbar riscos à saúde de pessoas que vivem em regiões silvestres, como zonas rurais, comunidades indígenas e regiões ribeirinhas.

Deste modo, uma solução no contexto de campanhas de saúde não deve negligenciar a diversidade de condições de vida no Brasil, mas deve considerar formas, meios, formatos, canais e estratégias para que a comunicação alcance a maior diversidade possível de públicos. A população deve encontrar informações que podem ser acessíveis a nível informacional (contra o uso apenas de linguagem escrita para pessoas da Comunidade Surda ou pessoas analfabetas). O tipo de canal utilizado também deve ser diverso, ao contrário de utilizar apenas aplicativo, considerando que há pessoas que não possuem dispositivos tecnológicos. Do mesmo modo, a diversidade de meios de informação deve ser possibilitada, evitando barreiras como campanhas de saúde veiculadas apenas por peças publicitárias que dependem da visão, não sendo acessíveis para pessoas cegas. Quando se trata da saúde, não prover acesso não significa apenas fazer com que pessoas percam uma informação, mas significa barrar e negligenciar o direito de promoção à saúde voltados para o bem-estar e qualidade de vida, assim como promover um ambiente com risco de ocorrência da desinformação.

Ao entrar em contato com o sistema de saúde, identificou-se vários desafios que giram em torno de uma falta de acolhimento e olhar humanizado para com os usuários do SUS, tanto a nível de tratamento quanto de espaços acolhedores. Isso reforça a necessidade de uma solução no contexto das campanhas pensar a comunicação encontrando a pessoa humana na sua dignidade em saúde, desde o acesso à informação até encontrar nos ambientes e pessoas da saúde um lugar de olhar e escuta humanizada, que olha para a pessoa por inteiro, suas relações, espaços sociais e família.

Para profissionais da APS, identificou-se principalmente dificuldades em operar os sistemas digitais do SUS. São muitos sistemas, que podem sofrer atualizações, o que demanda treinamento constante dos profissionais. Muitas operações são realizadas diariamente, o que indica a necessidade de os sistemas serem intuitivos, fáceis de usar e não introduzem barreiras adicionais para o trabalho dos profissionais de saúde. Para os gerentes da APS, por sua vez, identificou-se uma dificuldade de avaliar e monitorar a aplicação de campanhas, os recursos associados e o desempenho da equipe. Essa dificuldade é uma barreira para o gerenciamento da execução da campanha e na manutenção de indicadores de produção, como o número de domicílios visitados. Do mesmo modo, há dificuldade em registrar e sistematizar lições aprendidas de profissionais sobre uma campanha, conhecimentos estes que poderiam ser utilizados para beneficiar o planejamento e execução de campanhas futuras. Uma solução pode auxiliar pessoas gestoras a criarem uma cultura de planejamento e avaliação das ações, assim como de registro e valorização dos conhecimentos adquiridos pelos profissionais em aplicações futuras de campanhas.

---

[3] Último acesso em: 17/12/2024. Disponível em:
https://www.gov.br/saude/pt-br/assuntos/saude-de-a-a-z/f/febre-amarela

Para ACS e ACE, identificou-se problemas de atuação relacionados ao próprio trabalho no território. É desafiador conseguir atender uma população diversa e extensa, ainda mais considerando que cada pessoa pode ter situações de saúde específicas que demandam abordagens únicas de comunicação e cuidado. Ao atuar no território, agentes de saúde enfrentam dificuldades ambientais e fatores de risco (regiões epidêmicas, violentas, isoladas ou distantes), o que indica a importância de infraestrutura e material de trabalho (ex., protetor solar, EPIs). Para agentes, uma solução pode auxiliar a otimizar a atuação, por exemplo auxiliando a planejar o atendimento no território a fim de alcançar as famílias com comunicação e tempo de qualidade.

Também identificou-se para os agentes desafios únicos que se inserem em seu relacionamento com a população. Um problema central é a dificuldade em criar e manter vínculos com as famílias em um cenário extenso, dinâmico (famílias que se mudam) e de ter acesso aos domicílios (serem recebidos em visitas domiciliares). No processo de comunicação à saúde, agentes de saúde também têm o desafio de superar uma possível desinformação, onde a informação sendo comunicada pode entrar em conflito com, por exemplo, notícias falsas já circuladas entre a população e comportamentos derivados dessas notícias (ex., não querer vacinar-se). Neste contato com a população, há o desafio do não entendimento pela população da importância ou atribuição dos agentes de saúde, o que implica em um sentimento de falta de valorização profissional. Estes problemas revelam que uma solução no contexto de campanhas de saúde pode contribuir para fortalecer laços, vínculos, construção de entendimentos, significados e outros aspectos relacionais que são importantes na atuação de agentes de saúde juntamente à população.

## 4.3 Quadro de Prospecção de Valor

O Quadro de Prospecção de Valor (Ferrari et al., 2020) apresenta três questões para consolidar um entendimento de problema em termos de suas partes interessadas centrais, suas principais necessidades e dificuldades. Com a resposta das três questões, o quadro apresenta duas questões para permitir materializar os principais aspectos da solução: como ela pode melhorar a vida das pessoas, e como pode surpreender e inovar.

No Quadro, questionamos nossos conhecimentos sobre o domínio do problema que talvez já estejam fixos, nos perguntando sobre o quê realmente essa parte interessada precisa no contexto investigado. O quadro também permite estabelecer um cuidado explícito com o compromisso da solução em atender o valor das pessoas em suas necessidades e pensando em como a solução pode ser surpreendente e inovadora na vida das pessoas. A seguir, o Quadro 1 apresenta o preenchimento das 5 questões para o projeto de Campanhas da Atenção Primária à Saúde.

| | |
|---|---|
| **1. Quais são os stakeholders centrais?**<br>Agentes de Saúde (ACS/ACE) e Gestores Municipais da APS. | **4. Em quais aspectos a solução pode melhorar a vida dessas pessoas?**<br>Agentes: melhorar a efetividade na educação em saúde relacionada às campanhas, fazendo a mensagem chegar realmente à população.<br><br>Gestores Municipais da APS: ter melhor controle sobre as campanhas e entender indicadores relacionados à eficácia das campanhas. |
| **2. O que essas pessoas realmente precisam?**<br>Agentes: formas de facilitar seu trabalho na gestão do território e no relacionamento com as pessoas.<br><br>Gestores Municipais da APS: formas de acompanhar e facilitar a gestão das campanhas sendo aplicadas. | |
| **3. Quais dificuldades essas pessoas enfrentam?**<br>Agentes: dificuldade de aplicar a educação em saúde relacionada às campanhas.<br><br>Gestores Municipais da APS: complexidade no gerenciamento, aplicação e avaliação das campanhas de saúde. | **5. Como a solução pode surpreender e inovar?**<br>Agentes: fazendo com que o trabalho seja cada vez mais próximo da comunidade, reduzindo entraves no trabalho mesmo em cenários em que a população não sabe ou não valoriza sua atuação.<br><br>Gestores Municipais da APS: aumentando o apoio no planejamento, execução e acompanhamento das campanhas, assim como registro de aprendizados visando a evolução de campanhas futuras |

**Quadro 1.** Quadro de Prospecção de Valor. Modelo adaptado de Ferrari et al. (2020).

No quadro, retoma-se os agentes de saúde (incluídos ACS e ACE) e gestores municipais da APS como principais partes interessadas no contexto da campanha de saúde para esse projeto. Os agentes de saúde necessitam de formas de facilitar a gestão do território e no relacionamento com as pessoas, em um cenário em que precisam lidar com a demanda de muitos domicílios a serem visitados, e no qual a mensagem de saúde relacionada à campanha pode ficar em segundo plano ou passar despercebida pela população. Os gestores municipais da APS, por sua vez, necessitam de formas de acompanhar e facilitar a gestão das campanhas, uma vez que elas podem possuir indicadores associados que precisam ser cumpridos (ex: números de casas visitadas, quantidade de materiais distribuídos, etc.), que podem ser de difícil acompanhamento e avaliação.

Uma solução tecnológica de apoio ao gerenciamento e execução das campanhas de saúde pode melhorar a efetividade dos agentes de saúde em um de seus papeis principais, que é o de educação em saúde e bem-estar com a população do território. As campanhas, no geral, já possuem esse propósito de não focar na doença, mas na promoção da saúde como qualidade de vida. Os agentes de saúde, em seu contato com a população, são difusores de práticas, comportamentos e mensagens mais amplas de saúde voltada para a qualidade de vida e para o bem-estar físico, mental e social. Para os gestores municipais da APS, uma solução pode melhorar o trabalho de acompanhamento, principalmente da gestão e entendimento de indicadores que podem mostrar indícios de efetividade e oportunidades de melhorias na aplicação das campanhas.

A solução pode inovar ao fornecer funcionalidades que aproximem os agentes da população do território, facilitando seu trabalho, reduzindo entraves e processos manuais que afastam o agente da promoção à saúde, principalmente em cenários onde a população não conhece ou não valoriza a atuação destes agentes. Para gestores municipais da APS, uma solução pode inovar ao fornecer funcionalidades que apoiem e facilitem realizar o planejamento, execução e acompanhamento das campanhas, assim como registro de aprendizados dos envolvidos, possibilitando evolução em campanhas futuras.

### 4.4 Personas e Cenários de Interação

Com o preenchimento do Quadro de Prospecção de Valor, foram definidos cenários exemplificando as principais partes interessadas como *personas* no contexto da APS e sua interação com uma solução para gerenciamento de campanhas de saúde. Os cenários a seguir exemplificam as principais partes interessadas, as características situadas de seu contexto e suas e necessidades, assim como os desafios que surgem em sua maneira única de interagir com as campanhas.

**Persona 1. Lenir, mulher, 35 anos, Altamira (PA), Secretária Municipal de Saúde.**

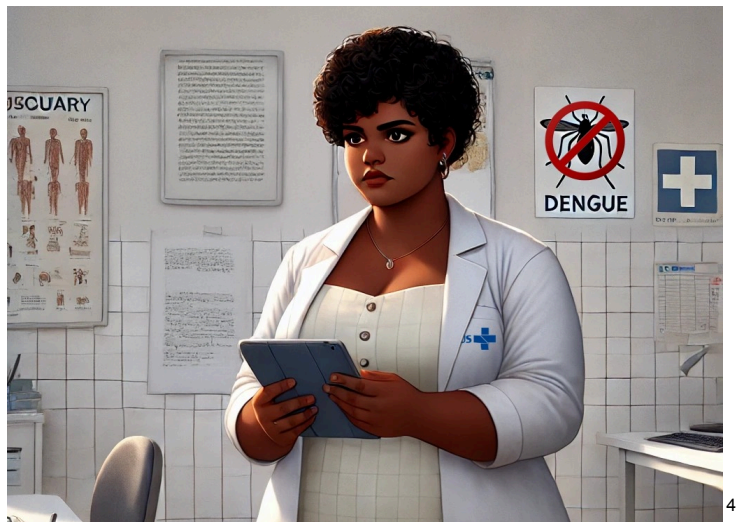

[4]

**Cenário 1. Lenir e o envio de notificações à Agentes Comunitários de Saúde.**
Lenir é uma mulher de 35 anos, que atualmente é Secretária de Saúde do Município de Altamira (PA). Ela costuma enfrentar dificuldades na comunicação com os Agentes Comunitários de Saúde (ACS) do seu município, que é o maior município do Brasil e o terceiro maior do mundo em extensão territorial e consequentemente enfrenta diversas desigualdades, incluindo dificuldade de acesso à internet e dispositivos eletrônicos. O município de Altamira foi beneficiado pelo Projeto de Lei 7079/17, que determina o fornecimento de equipamentos como tablets e smartphones para o trabalho de ACS, ACE e Agentes de Ação Social (AAS). Esses profissionais utilizam o aplicativo e-SUS Território nos tablets para registro e atualização do Prontuário Eletrônico do Cidadão (PEC), de terrenos, famílias e cidadãos. Mesmo com a dificuldade de comunicação e conectividade, Lenir costuma enviar mensagens no grupo de um aplicativo de troca de mensagens. Foi assim que ela mandou uma mensagem para avisar as pessoas do grupo sobre a campanha de cuidados constantes na prevenção à dengue. João, um ACS de 50 anos que não utiliza aplicativo de troca de mensagens e utiliza o tablet fornecido pelo Ministério da Saúde em suas visitas domiciliares não recebeu a mensagem, assim como outros ACS que não têm esse aplicativo e, mesmo os que têm nem sempre leem as mensagens. Lenir gostaria de poder enviar mensagens para sua equipe de ACS por meio de notificações no aplicativo e-SUS Território, mas não existe essa possibilidade.

---

[4] Imagem produzida pela ferramenta de imagem generativa (DALL-E 3) disponível no chatgpt.

Este primeiro cenário descreve os desafios de um gestor da APS, como um secretário de saúde, que possui desafios no gerenciamento das campanhas e na comunicação com os agentes de saúde. No cenário, foi prospectada uma solução que apoia o gerenciamento de campanhas e com funcionalidades que facilitam a comunicação com os agentes na cidade de Altamira (PA), uma das maiores em extensão territorial e com contextos de vulnerabilidade socioeconômica.

**Persona 2. Charles, homem, 29 anos, Curitiba (PR), Agente de Combate a Endemias.**

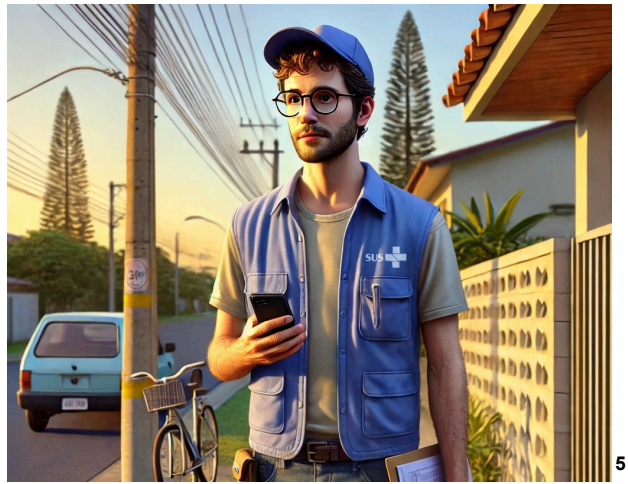

[5]

**Cenário 2. Charles e a dificuldade de receber informações do Secretário Municipal de Saúde.**
Charles é um homem de 29 anos, que trabalha como Agente de Combate a Endemias (ACE) no município de Curitiba (PR), que possui uma grande concentração populacional (população estimada em 1,8 milhão de habitantes em 2024), um fato comum nas capitais. Com o aumento na proliferação do mosquito *aedes aegypti* e consequente aumento no número de casos de dengue no município, Charles precisa visitar as residências no seu dia a dia em busca de focos do mosquito e para fazer campanhas de conscientização e prevenção. No entanto, ele costuma enfrentar dificuldade em ser recebido nessas visitas, principalmente em condomínios fechados. Essa dificuldade se deve a diversos motivos, como o fato de haver golpes em que as pessoas se passam por profissionais para serem recebidas nas residências e cometerem crimes, ou de não haver pessoas nas residências no horário das visitas. Charles tem um grupo em um aplicativo de mensagens com moradores das residências de sua área de visitas, no qual costuma mandar mensagens e tenta agendar essas visitas para que consiga encontrar os moradores em casa e ser recebido. Contudo, nem todos os moradores estão nesse grupo e mesmo os que estão nem sempre respondem suas mensagens. Deste modo, às vezes Charles tem que trabalhar fora do horário formal para encontrar as pessoas em suas casas (como na hora do jantar). Charles gostaria que a comunicação com os domicílios do seu território fosse facilitada. Que houvesse, por exemplo, a possibilidade de agendamento e uma notificação oficial da Secretaria Municipal de Saúde no aplicativo do SUS, indicando que haveria uma visita do ACE.

---
[5] Imagem produzida pela ferramenta de imagem generativa (DALL-E 3) disponível no chatgpt.

O segundo cenário apresenta desafios de um agente de combate à endemias em Curitiba (PR), uma capital que oferece barreiras para visitas presenciais em domicílios, o que impede o combate efetivo à dengue com ações mais efetivas. O cenário prospecta funcionalidades para uma solução que apoia na comunicação e agendamento de visitas para favorecer o trabalho de vigilância à saúde.

**Persona 3. Márcia, mulher, 47 anos, Teresina (PI), Agente Comunitária de Saúde.**

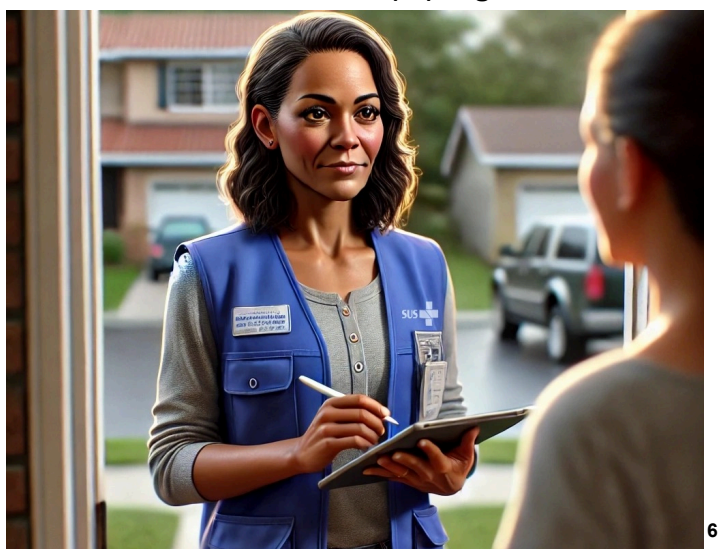
[6]

**Cenário 3. Márcia, os desafios de criar vínculos com as famílias e em conhecer lições aprendidas da aplicação de campanhas de saúde.**

Márcia é uma Agente Comunitária de Saúde no município de Teresina, no Piauí. Ela enfrenta desafios na sua atuação como Agente na capital, pois não consegue aprofundar seu vínculo com as famílias. Márcia precisa trafegar longas distâncias, nem sempre encontrando as pessoas em seus domicílios. Quando encontra alguém em casa, Márcia precisa atualizar muitos dados do domicílio, visto que em capitais muitas pessoas mudam de endereço frequentemente. Como boa parte do tempo é utilizado para o cadastro ou atualização dos dados, Márcia fica com pouco tempo para criar vínculos com a família e promover um processo de escuta e diálogo sobre a promoção de campanhas de saúde mais amplas relacionadas ao bem-estar físico, mental e social. Márcia gostaria que o aplicativo e-SUS Território a ajudasse a registrar e relembrar informações sobre as famílias que lhe permitisse aprofundar os vínculos e uma compreensão mais ampla das famílias.

---

[6] Imagem produzida pela ferramenta de imagem generativa (DALL-E 3) disponível no chatgpt.

Márcia gostaria, por exemplo, de lembrar dos apelidos carinhosos que a comunidade tem para algumas pessoas da família, assim como os hobbies e espaços que essas pessoas frequentam. Márcia entende que relembrar essas informações permitirá criar maior confiança entre ela e a comunidade, assim como entender as estratégias de comunicação de saúde associada com os espaços sociais a qual as pessoas do território frequentam e vivem. Márcia também gostaria de ter um espaço onde pudesse compartilhar e dialogar com outros Agentes sobre boas práticas, lições aprendidas e demais informações importantes sobre seu território em campanhas de saúde. Muitas vezes, as informações sobre o planejamento e execução do próprio território (e de campanhas de cidades vizinhas que compartilham a mesma situação social) são esquecidas ou perdidas. Com isso, a cada campanha não se aproveita adequadamente os conhecimentos e lições aprendidas que os agentes produzem em sua atuação. Márcia gostaria de compartilhar suas boas práticas, assim como conhecer boas práticas de outras regiões de seu estado.

Por fim, o terceiro cenário apresenta uma agente comunitária de saúde na cidade de Teresina (PI) e os desafios que ela possui na criação de vínculos com as famílias de seu território. O cenário prospecta funcionalidades que permitem a agente de saúde ter um contato mais próximo e humano com as famílias, em direção a uma atuação mais humana e consciente.

As *personas* e cenários ilustram realidades de diferentes partes interessadas centrais no contexto das campanhas de saúde, tanto da gerência (Secretária de Saúde), quanto dos executores das campanhas com as famílias (ACS e ACE). As partes interessadas são provenientes de diversas regiões do país e encontram desafios situados de comunicação e execução das campanhas. Em cenários rurais, por exemplo, há desafios de infraestrutura e acesso. Cenários urbanos, por sua vez, apresentam desafios de comunicação e de manutenção de vínculos entre agentes e as famílias em territórios vivos e dinâmicos. A definição deste cenários informa a abrangência de casos que uma solução de APS precisa atender e apoia a identificação de requisitos para a solução, apresentados na subseção a seguir.

## 4.5 Requisitos Prospectivos da Solução no Framework Semiótico

O processo de entendimento de problema, as personas e os cenários foram informações de entrada para a definição de requisitos que especificam uma solução no contexto de criação e gerenciamento de campanhas de saúde na APS. Os requisitos estão apresentados no no formato de histórias de usuário (HU) no Quadro 2, sendo referentes aos aspectos humanos e sociais da solução.

Os requisitos do nível do Mundo Social representam aspectos do mundo que as pessoas vivem que não podem ser ignorados e que definem aspectos ou funcionalidades para a solução. Foram identificados requisitos que representam a importância do sistema ser acessível para a maior diversidade de condições biológicas e socioeconômicas possíveis (HU1) em um contexto de soberania tecnológica brasileira (HU2). Em termos dos relacionamentos no entorno das Campanhas, a solução pode beneficiar e auxiliar no aprofundamento dos relacionamentos entre as partes interessadas da APS e a população (HU4 - HU7).

> *Mundo Social: Valores, Cultura, Crenças, Relações e Impacto Social da Solução*
>
> HU1. Eu, como uma pessoa usuária do SUS, QUERO que minha diversidade de condições sensório-motoras, mentais, socioculturais, e socioeconômicas sejam consideradas PARA QUE eu não tenha barreiras de acesso aos serviços e comunicação em saúde.
>
> HU2: Eu, como o ministério da saúde QUERO que o sistema seja desenvolvido utilizando tecnologias abertas e livres PARA QUE a soberania nacional seja protegida e a LGPD seja seguida.
>
> HU3. Eu, como um agente de saúde QUERO que o sistema auxilie a intermediar o contato e relacionamento com meus colegas de trabalho e a população do meu território PARA QUE o trabalho de comunicação seja facilitado e os desafios de comunicação sejam superados.
>
> HU4. Eu, como um agente de saúde QUERO que o sistema não provoque maior dificuldade ou impedimento PARA QUE o propósito da educação em saúde das campanhas não seja impedido.
>
> HU5. Eu, como uma pessoa gestora da APS QUERO que o sistema possibilite compartilhar práticas, saberes e experiências entre os agentes (ACS e ACE) PARA promover a integralidade da ação e dos profissionais na saúde.
>
> HU6. Eu, como um agente de saúde QUERO que o sistema possua um feed no qual agentes do Brasil inteiro publiquem as maneiras como estão divulgando e operacionalizando as campanhas em seus territórios PARA QUE o conhecimento de boas práticas e lições aprendidas seja compartilhado e melhore minha prática.
>
> HU7. Eu, como uma pessoa gestora da APS QUERO que o sistema possua funcionalidades sociais entre agentes de saúde e população do território PARA QUE se fortaleça a interação e vínculo entre agentes e comunidades.
>
> HU8. Eu, como um gestor da APS QUERO que o sistema facilite o acolhimento dos usuários que precisam sair da APS e receber o Atendimento Especializado Ambulatorial, por exemplo indicando exatamente quais profissionais da ESF devem ser consultados e tendo uma interligação com o sistema de agendamento de consultas PARA QUE a atenção primária possa encaminhar corretamente os usuários nos demais níveis de atenção à saúde e promover a integralidade do cuidado.
>
> HU9. Eu, como o C3SL QUERO que o sistema respeite a LGPD ao coletar, tratar, processar e armazenar dados dos usuários do SUS PARA QUE a legislação seja respeitada e promova maior segurança, privacidade e confiança no uso do sistema.

**Quadro 2.** Requisitos referentes a aspectos sociais e humanos da solução.

Requisitos do nível Pragmático emergem a partir das intenções, desejos e expectativas das partes interessadas. Foram identificados requisitos que representam expectativas de secretárias de saúde e gerentes da APS, como avaliar e acompanhar a execução das campanhas (HU10). Pela perspectiva de agentes de saúde, foram identificados requisitos que apoiam no trabalho da execução da campanha (HU11) e no relacionamento com a população do respectivo território, facilitando a comunicação pró-ativa (HU12) e um acompanhamento mais amplo de saúde das famílias (HU13, HU14, HU15).

> *Pragmático: Intenções das Partes Interessadas para a Solução*
>
> HU10. Eu, como uma pessoa gestora da APS ou um agente de saúde QUERO que o sistema facilite meu entendimento e acompanhamento de indicadores relacionados às Campanhas de Saúde PARA QUE eu possa compreender a execução e atuar para o prol do sucesso da campanha.
>
> HU11. Eu, como uma pessoa gestora da APS ou um agente de saúde QUERO que o sistema agrupe as informações sobre as Campanhas de Saúde em um único local, com fácil acesso (ex: materiais de divulgação, orientações, *links* e vídeos) PARA facilitar o acesso e compartilhamento com a população de materiais importantes na comunicação em saúde.
>
> HU12. Eu, como um agente de saúde QUERO que o sistema promova feedback do usuário de forma proativa, onde são os próprios usuários do SUS que informam para o ACE/ESF sobre aspectos importantes de saúde, como gravidez, nascimento de criança e emergências PARA QUE o processo de comunicação com a população seja bidirecional, agilizando no levantamento de demandas e conhecimento sobre situações de risco.
>
> HU13. Eu, como um agente de saúde QUERO que o sistema apoie a registrar fatores socioeconômicos (ex., desigualdade de renda, dependência química, nível alto de violência) e geográficos (ex., deslizamentos e inundações) de risco à saúde que uma família em particular possua PARA QUE eu possa reconhecer facilmente fatores de risco que as famílias possuem em meus atendimentos e acompanhar essas famílias mais de perto.
>
> HU14. Eu, como uma pessoa gestora da APS QUERO que o sistema mostre um mapa do território com quais famílias possuem fatores socioeconômicos e geográficos relacionados a uma determinada Campanha PARA favorecer o planejamento das visitas domiciliares e focar a atenção da campanha em seus públicos-alvos naquele território.
>
> HU15. Eu, como um agente de saúde QUERO que o sistema facilite o acompanhamento das informações relativas às pessoas com condições de saúde crônicas e vulneráveis (ex., acamados) PARA fornecer maior atenção para populações que demandam maior cuidado.

**Quadro 3.** Requisitos referentes a aspectos pragmáticos da solução.

Requisitos do nível Semântico (Quadro 4) emergem dos significados e sentidos que as pessoas produzem em seu contexto. Foram identificados requisitos referentes à produção e registro dos conhecimentos (HU20) que Agentes adquirem na execução das campanhas, que podem ser aproveitados como lições para as campanhas seguintes (HU18).

Requisitos para favorecer a comunicação na ESF e entre agentes e a população também foram identificados (HU19 e HU21). Por fim, foi identificado um requisito que beneficia gerentes a entenderem de forma assertiva a situação de execução de uma campanha (HU17) e um requisito que define para a solução o uso de linguagem situada e regional para favorecer a acessibilidade e compreensão da mensagem (HU16).

> *Semântico: Compreensão e Significados em torno da Solução*
>
> HU16. Eu, como um agente de saúde QUERO que a comunicação em saúde use linguagens e significados regionais ao entrar em contato com uma pessoa PARA possibilitar maior entendimento da mensagem de saúde relacionada à campanha.
>
> HU17. Eu, como uma pessoa gestora da APS QUERO que o sistema apresente indicadores de Campanhas de Saúde previstas, finalizadas, em andamento, e os respectivos números associados às campanhas PARA facilitar o acompanhamento e avaliação das campanhas
>
> HU18. Eu, como um agente de saúde QUERO que o sistema permita fazer comentários sobre as Campanhas em Andamento PARA angariar feedback, opiniões, sugestões de melhoria e refinamentos para o planejamento da próxima campanha.
>
> HU19. Eu, como um agente de saúde QUERO que o sistema tenha canais bidirecionais de comunicação compartilhada entre ACS e ACE de minha eSF PARA que ambos os profissionais sejam capazes de entender o que cada um está fazendo ou fez em um determinado período, evitando retrabalho entre os agentes.
>
> HU20. Eu, como um agente de saúde QUERO que o sistema registre dúvidas que a população comunica comigo, mas que eu ainda não sei responder, principalmente em relação às novas arboviroses e situações epidêmicas PARA ter uma comunicação mais assertiva com a população no surgimento de dúvidas, para armazenar o conhecimento importante na resposta das dúvidas da população e para evitar a desinformação.
>
> HU21. Eu, como um agente de saúde QUERO que o sistema armazene informações do meu relacionamento interpessoal com usuários do SUS, por exemplo informações que humanizem o usuário, como o "apelido" da pessoa, quais espaços ela frequenta (ex., igreja, grupo de prática local), atividades ou hobbies essa pessoa tem, principalmente no contexto de territórios com *turnover* alto de agentes de saúde PARA facilitar o conhecimento da população de forma a estabelecer uma proximidade e criar vínculos.

**Quadro 4.** Requisitos referentes a aspectos semânticos da solução.

Requisitos do nível Sintático (Quadro 5) definem aspectos relacionados à estrutura, forma e formatos de uma solução. Foram identificados requisitos relacionados à diversidade de formas de apresentação da informação (HU22, HU23, HU26 e HU27). Também foram identificados requisitos que definem a possibilidade de edição dos materiais informativos de campanha (HU24) e associar indicadores da campanha com a visão do território (HU25).

> *Sintático: Formas, Estruturas, Padrões e Lógica da Solução*
>
> HU22. Eu, como um agente de saúde ou um usuário do SUS QUERO que o sistema disponibilize diversas formas de apresentar a informação (visual, auditiva, motora, libras) relacionada às Campanhas de Saúde PARA que eu possa compreender a informação.
>
> HU23. Eu, como um agente de saúde QUERO que compartilhar o conteúdo das Campanhas de Saúde por redes sociais (ex: WhatsApp, Telegram, Facebook, TikTok, etc.) PARA favorecer a disseminação da mensagem de saúde para a população que tenho conectada em minhas redes.
>
> HU24. Eu, como um agente de saúde QUERO editar ou acessar os modelos editáveis da Campanha PARA editar e refinar os materiais para a minha realidade local.
>
> HU25. Eu, como uma pessoa gestora da APS QUERO associar indicadores da campanha com dados quantitativos da microárea/população, por exemplo em uma campanha para gestantes, o sistema pode apresentar dados referentes às famílias/domicílios de gestantes PARA ajudar a planejar a execução da campanha com pontos de atenção do território.
>
> HU26. Eu, como uma pessoa gestora da APS QUERO que o sistema permita anexar materiais de divulgação e informação da Campanha na forma de imagens, PDFs, vídeos com legenda e LIBRAS e áudio PARA que a comunicação em saúde alcance a população independente de suas condições.
>
> HU27. Eu, como o C3SL QUERO que o sistema seja desenvolvido seguindo padrões de acessibilidade na web (WCAG) e o padrão de acessibilidade do governo (eMAG) PARA que o sistema seja acessível para a diversidade de condições e necessidades dos mais de 400 mil Agentes de Saúde no Brasil.

**Quadro 5.** Requisitos referentes a aspectos sintáticos da solução.

Requisitos do nível Empírico definem características de frequência e capacidade relacionados à transmissão de informação e dados. Foi definido que a solução deve utilizar vários canais para disseminar a informação (HU29), que deve seguir os padrões de transmissão e interoperabilidade de dados do governo e com sistemas de informação do governo (HU28 e HU30).

Por fim, foi definido que o sistema deve funcionar com modo offline, com sincronização quando o dispositivo estiver com conexão wi-fi, visto a atuação de Agentes em contextos com infraestrutura tecnológica diversa (HU31). As histórias de usuário referentes aos aspectos empíricos da Solução são apresentadas no Quadro 6 a seguir.

---

*Empírico: Frequência, Capacidades, Redundância, Eficiência da Solução*

HU28. Eu, como o C3SL ou o Ministério da Saúde QUERO que o sistema seja desenvolvido seguindo padrões de interoperabilidade do Governo Brasileiro (ePING) PARA que o sistema possa comunicar e compartilhar dados com os demais sistemas de informação em saúde.

HU29. Eu, como uma pessoa gestora da APS QUERO veicular uma mensagem por diversos canais (ex., SMS, ligação telefônica, aplicativo, email, etc.) PARA que a comunicação com os agentes de saúde ocorra independente das condições tecnológicas de acesso.

HU30. Eu, como o C3SL ou o Ministério da Saúde QUERO que o sistema seja desenvolvido planejando sua integração com a RNDS e a interoperabilidade dos dados obtidos via sistema com outros sistemas-alvo de interesse do projeto PARA que o sistema possa integrar esforços para a Estratégia de Saúde Digital para o Brasil.

HU31. Eu, como um agente de saúde QUERO que o sistema tenha um modo offline e sincronização dos dados quando estiver conectado em rede Wi-Fi PARA que eu possa utilizar o sistema mesmo em condições sem acesso ou com acesso limitado à internet.

---

**Quadro 6.** Requisitos referentes a aspectos empíricos da solução.

Por fim, no Mundo Físico são definidos aspectos de infraestrutura tecnológica e física da solução. Na perspectiva de soberania nacional, segurança e privacidade das pessoas, foi definido que a solução deveria funcionar e ser instalada em território brasileiro, em servidores de dados nacionais (HU32).

Considerando contextos em que Agentes de Saúde não tenham acesso a dispositivos tecnológicos para realizar atualizações com relação à campanha, foi definido que é necessário possibilitar que os dados sejam inseridos posteriormente, por exemplo no ambiente físico de uma UBS (HU33). Por fim, é preciso conceber uma solução capaz de operar em dispositivos de baixo poder de processamento e armazenamento (HU34), tendo compatibilidade com tecnologias assistivas (HU35). As histórias de usuário deste nível do Framework Semiótico são apresentadas no Quadro 7 a seguir.

> *Mundo Físico: Hardware e Infraestrutura da Solução*
>
> HU32. Eu, como o C3SL/Ministério da Saúde QUERO que a infraestrutura de armazenamento (banco de dados, servidor e outros elementos de infraestrutura) seja viabilizada em território brasileiro PARA respeitar a soberania nacional, assim como a segurança e privacidade de dados dos cidadãos brasileiros.
>
> HU33. Eu, como uma pessoa gestora da APS QUERO que o sistema possua uma opção de interação para pessoas sem dispositivos tecnológicos pessoais PARA permitir a inserção dos dados de visitas domiciliares mesmo quando um dispositivo tecnológico não esteja disponível.
>
> HU34. Eu, como um agente de saúde QUERO que o sistema funcione no dispositivo fornecidos a mim pelo governo (ex., tablets) ou no meu dispositivo PARA que o sistema possa ser utilizado com dispositivos da minha realidade, como materiais de trabalho e celulares populares (até 2GB de RAM).
>
> HU35. Eu, como um agente de saúde com deficiência QUERO que o sistema seja compatível com tecnologias assistivas PARA que eu possa utilizá-lo independente de minhas condições.

**Quadro 7.** Requisitos referentes a aspectos de infraestrutura física da solução.

Com o uso de histórias de usuário, as partes interessadas que originaram cada requisito são mantidas. Do mesmo modo, as justificativas ou contextos motivadores da especificação do requisito estão associados. Os requisitos são prospectivos e definem funcionalidades, restrições, aspectos de interface e apresentação, disponibilidade de recursos e infraestrutura para a solução. A partir desses requisitos, foram definidos protótipos da solução na perspectiva do cadastro e gerência das campanhas.

## 4.6 Protótipos de Média Fidelidade

A solução foi materializada por meio de protótipos navegáveis de média fidelidade desenvolvidos com o aplicativo Adobe XD. Nesta prototipação, a maior preocupação não foi fornecer fidelidade em termos estéticos, mas exemplificar interfaces com suas funcionalidades e fluxos de interação. Na apresentação das Figuras 1-7, retomamos algumas das histórias de usuário que foram materializadas visualmente nos protótipos.

Não foi o objetivo desta etapa da pesquisa materializar todas as histórias de usuário de forma exaustiva. Em uma perspectiva iterativa-incremental, planejou-se a prototipação de uma porção do sistema na perspectiva de pessoas gestoras da APS, devido a definição desta prioridade para o projeto oriunda do MS. Em etapas futuras da pesquisa, serão produzidos protótipos do ponto de vista de agentes de saúde e outras partes interessadas.

O processo de prototipação utilizou como material de entrada o processo de Design Socialmente Consciente desenvolvido, no qual o entendimento do problema e as histórias de usuário apoiaram a contextualizar e definir os elementos a serem prototipados, a materializar a apresentação e fluxos de interação. Uma análise do estilo informacional e de formatação de websites do Governo Federal foi realizada de modo que os protótipos estivessem alinhados com o estilo de componentes já utilizado pelo Governo. Com isso, tem-se uma proposta inicial de uma solução para gerenciar a publicação e execução de campanhas de saúde.

A Figura 2 representa a tela inicial do Gerenciador de Campanhas. As principais funcionalidades são relacionadas ao próprio gerenciamento das campanhas e estão apresentadas ao centro com ícones para acesso rápido: Autorizar, Cadastrar, Pesquisar e Retomar Rascunho de Campanhas, assim como a funcionalidade de Ver Relatórios sobre as Campanhas. Essa visão é voltada para o Secretário de Saúde ou quem gerencia a APS e as respectivas campanhas de saúde.

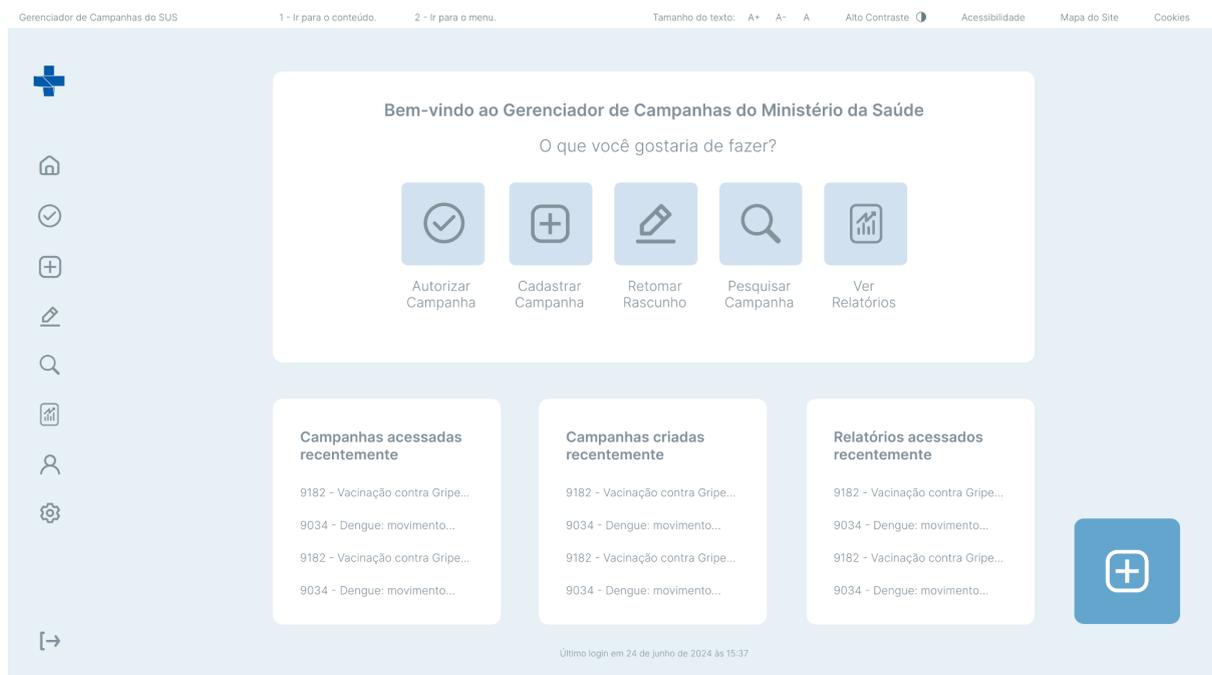

**Figura 2.** Página Inicial do Gerenciador de Campanhas de Saúde.

A Figura 2 ainda apresenta um menu superior, com opções de acessibilidade (HU22), assim como um menu lateral, com algumas opções de configuração e perfil, por exemplo. Atalhos para campanhas ou relatório de campanhas acessados recentemente estão disponíveis, visando apresentar indicadores de execução de campanhas como especificado na HU10 e HU17. Por fim, no canto inferior direito há um grande botão voltado para a criação rápida de campanhas. Esse design privilegia o acesso rápido e a simplicidade no acesso, imaginando um cenário de trabalho em uma UBS que demande assertividade e clareza na interação.

Ao acessar a funcionalidade de "Cadastrar Campanha", foi definido e prototipado um fluxo de telas voltado para o cadastramento das campanhas de saúde, dividindo a complexidade de informações associadas à uma campanha em seis passos. A Figura 3 representa a tela inicial, indicando do lado esquerdo números de um a seis que representam os passos do cadastro.

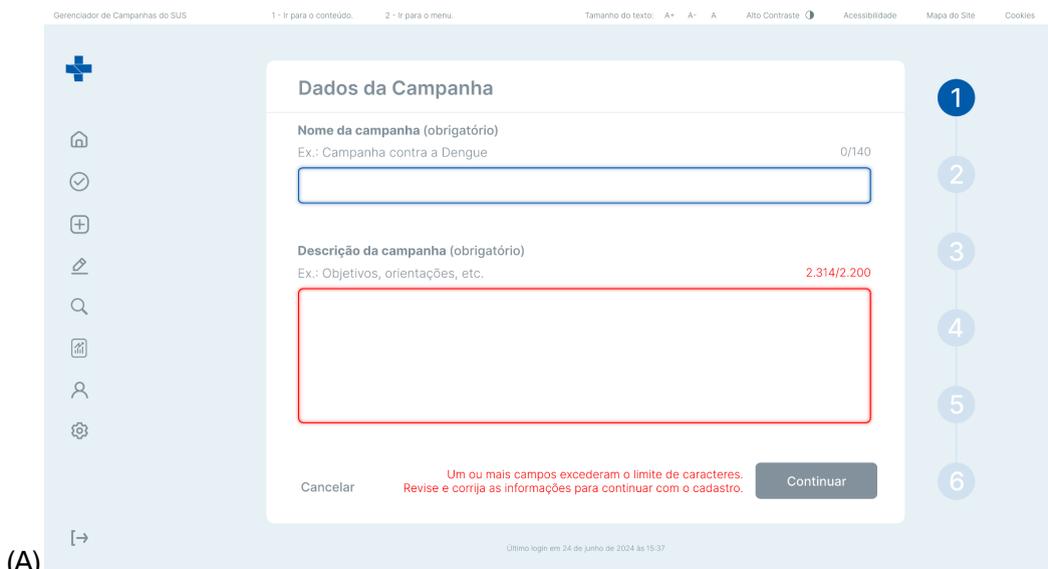

(A)

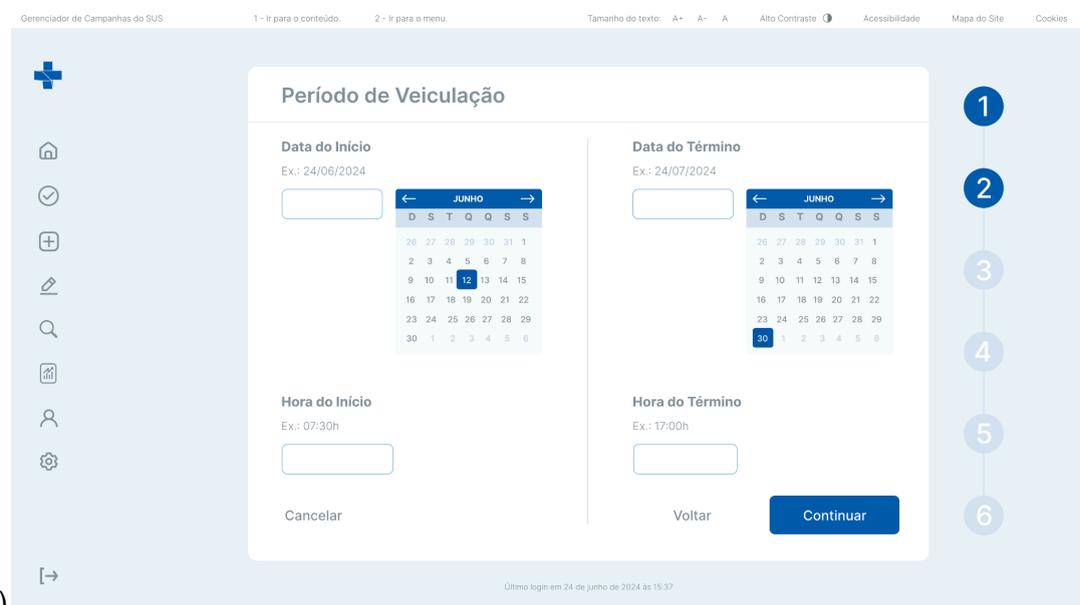

(B)

**Figura 3.** Página de Cadastro de Campanhas de Saúde. Figura 3-A apresenta a tela principal, voltada para cadastro do nome e descrição da campanha. A Figura 3-B, por sua vez, apresenta a data de início e término da campanha.

Na primeira tela de Cadastro (Figura 3-A) aparecem o cadastro do nome e descrição da campanha, com apresentação de exemplos. A segunda tela é voltada para o cadastro dos prazos de início e término da campanha, informações importantes para a APS ao gerenciar o calendário de campanhas de saúde, por exemplo na existência de campanhas a nível local (ex., campanha contra um surto de arbovirose local). que podem ocorrer de forma paralela a campanhas de nível nacional (ex., vacinação da gripe).

A terceira tela de Cadastro, apresentada na Figura 4, representa a seleção de informações sobre o público-alvo da campanha de saúde em questão. Algumas opções são indicadas, para facilitar o preenchimento, por exemplo referente à faixa etária do público-alvo, assim como o gênero. Informações mais detalhadas sobre o público-alvo podem ser inseridas na seção "Características Específicas", que apresenta um campo para cadastrar palavras-chave que representam o público-alvo.

**Figura 4.** Página de Preenchimento de Características do Público-Alvo das Campanhas

      Essa noção de características específicas tem o propósito de estar alinhado com o conceito de pluralidade, entendido como um movimento para a ampliação de intersecções entre conceitos como diversidade, equidade, acessibilidade e inclusão (de Oliveira et al., 2024). Prestar atenção às características da população permite trazer à tona a diversidade de condições e clivagens sociais que afetam o processo de saúde doença dos indivíduos, das famílias e coletividades, como indicado na HU01.

      A Figura 5 a seguir representa o cadastro das áreas e microáreas associadas à Campanha. Essa funcionalidade visa proporcionar flexibilidade no momento da criação da Campanha, pois ao mesmo tempo que permite criar campanhas a nível nacional (pelo próprio Ministério da Saúde), também permite criar ações de promoção à saúde cada vez mais situadas em realidades socioeconômicas e culturais locais, atendendo a demandas de promoção à saúde específicas para uma determinada região, como especificado na HU14.

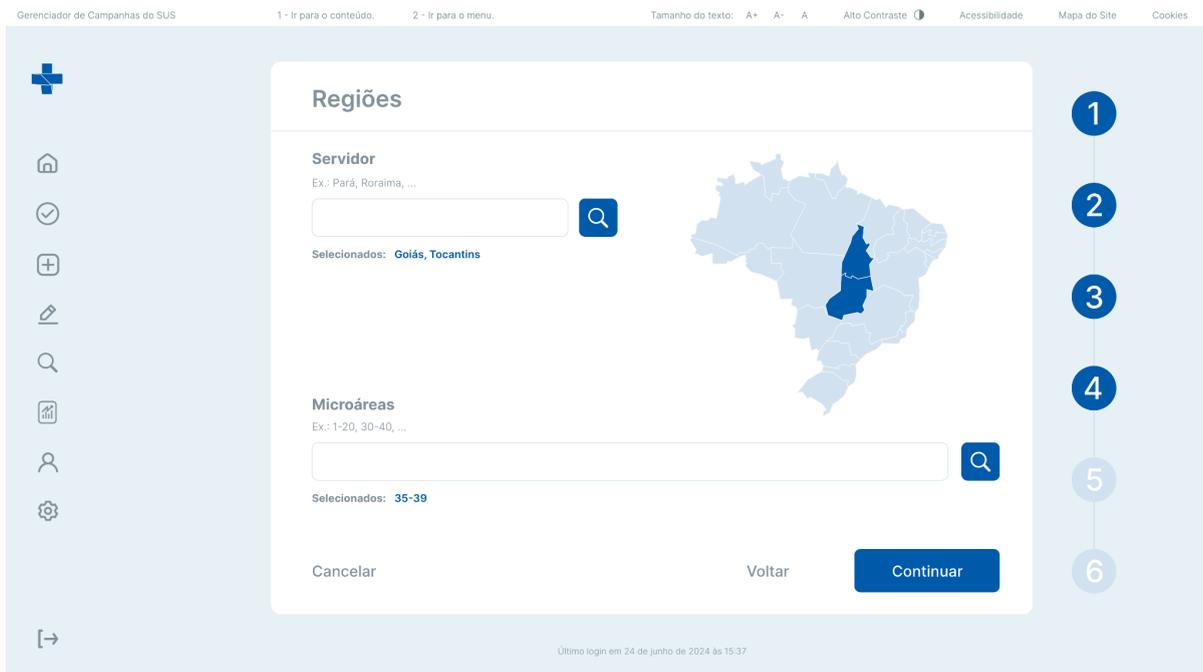

**Figura 5.** Página de Preenchimento de Regiões e Microáreas Associadas à Campanha.

Com essa funcionalidade, gestores podem ter uma visão global ou específica das campanhas de saúde em sua região. A nível estadual ou federal, por exemplo, é possível pensar e acompanhar campanhas realizadas em realidades compartilhadas entre municípios ou estados diferentes, por exemplo em fronteiras intermunicipais e interestaduais. A nível municipal, é possível planejar campanhas e ações de promoção à saúde que fazem sentido entre um subconjunto de microrregiões do território.

A Figura 6 apresenta a configuração da modalidade de envio da campanha, assim como os materiais informacionais associados.

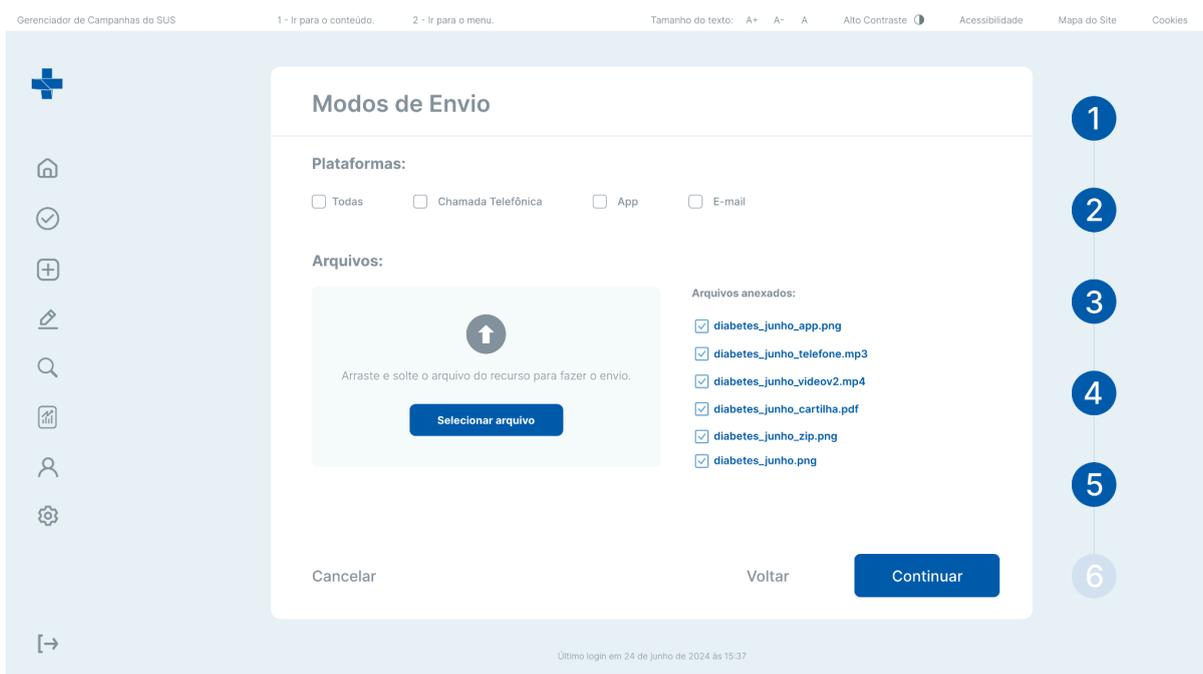

**Figura 6.** Página de Preenchimento das Modalidades de Envio e Materiais Associados à Campanha.

Em 2024, estima-se que 402.777 agentes de saúde estão atuando na APS brasileira (Rodrigues, 2024). Pensando na diversidade de Agentes de Saúde que atuam em diversas realidades do Brasil, incluindo em regiões rurais, ribeirinhas e quilombolas, é preciso disponibilizar acesso na maior diversidade de meios possíveis. Esta página representada na Figura 6 permite então enviar a mensagem das campanhas por meio de SMS, Chamada Telefônica, por meio de aplicativo e por meio de E-mail, funcionalidade que foi especificada na HU29.

A funcionalidade de "Arquivos", especificada na HU26, visa associar à campanha materiais informativos, como panfletos, banners, áudios, vídeos, arquivos PDF, dentre outros tipos de arquivos que podem favorecer o trabalho de agentes de saúde. A diversidade de arquivos é relevante para a mensagem ser acessível para pessoas com deficiência, analfabetas ou sem acesso a internet ou a dispositivos tecnológicos. Esse tipo de material é um auxílio importante no trabalho de promoção e vigilância à saúde, embora devam ser acessíveis para a diversidade populacional e refletir linguagens locais (Brito et al., 2020). Deste modo, é importante que os arquivos associados à uma campanha sejam, na medida do possível, editáveis posteriormente pelos Agentes de Saúde em suas realidades locais (como especificado na HU24), mudando por exemplo a linguagem utilizada para refletir termos, sentidos e significados regionais.

Por fim, a Figura 7 apresenta uma página que revisa todos os dados da Campanha antes de cadastrar a Campanha. O propósito é minimizar os erros no momento de cadastro da campanha, já que são muitos dados associados, como especificado na HU11.

**Figura 7.** Página de Revisão dos Dados da Campanha sendo Cadastrada.

Caso os dados estejam corretos, a campanha pode ser cadastrada. Caso não, o usuário pode voltar nas etapas anteriores para realizar as modificações necessárias. Essa prototipação teve como foco os usuários que gerenciam as Campanhas de Saúde, como gerentes da APS ou Secretários de Saúde. Nos próximos passos dessa pesquisa, ocorrerá o refinamento e aprofundamento deste protótipo de média fidelidade, assim como a validação das telas propostas com demais profissionais do grupo de pesquisa e desenvolvimento da Universidade Federal do Paraná (UFPR) e do MS.

Também busca-se prototipar e avaliar a solução pela perspectiva de Agentes de Saúde (ACS e ACE), materializando as funcionalidades do recebimento da campanha pelos agentes e a execução da campanha no território. Para isso, serão realizadas atividades envolvendo representantes do público-alvo. Um projeto de pesquisa foi elaborado e está sendo submetido ao Comitê de Ética da UFPR para viabilizar a condução de atividades com o público-alvo da solução.

## 5. Considerações Finais

Este relatório apresenta os resultados de uma investigação sobre como apoiar o gerenciamento de campanhas da APS, uma estratégia importante na difusão do direito à saúde para a população brasileira. De acordo com o framework do Design Socialmente Consciente, o design começa com o entendimento e definição do contexto ou problema a ser investigado (Baranauskas et al., 2024). Nesta pesquisa, o propósito foi entender o contexto abrangente, de modo que a própria definição deste contexto leve em conta os valores básicos de uma saúde pública e acessível, assim como a diversidade cultural de desafios socioeconômicos, culturais e técnicos que são singulares ao contexto brasileiro.

Como principais resultados, é possível explicitar que há dezenas de partes interessadas que são responsáveis pela estratégia de comunicação em saúde e que são foco desta comunicação. As partes interessadas são diversas e possuem condições geográficas, culturais e socioeconômicas igualmente únicas, que foram consideradas para a identificação de desafios, dificuldades e barreiras no Quadro de Avaliação. Com base nessa identificação, o Quadro de Prospecção de Valor nos possibilitou ratificar os ACS, ACE e Gestores Municipais da APS como principais partes interessadas no contexto das campanhas da APS e consolidar o entendimento dos problemas e barreiras enfrentados por eles. Os Agentes de Saúde, por exemplo, enfrentam desafios na gestão do território e no relacionamento com a população, muitas vezes sendo sobrecarregados pela demanda de visitas domiciliares, o que pode prejudicar a eficácia das campanhas. Já os Gestores Municipais da APS lidam com a complexidade do acompanhamento de indicadores, como número de visitas e materiais distribuídos, tornando a gestão difícil de ser monitorada.

Tendo em vista as principais partes interessadas no gerenciamento de campanhas da APS, foram criadas *personas* e cenários para representar as especificidades de cada uma, destacando suas interações, necessidades e desafios que podem surgir a partir delas. Nos cenários rurais, por exemplo, os principais desafios envolvem infraestrutura e acesso, enquanto em cenários urbanos, as dificuldades estão relacionadas à comunicação e à manutenção de vínculos entre agentes e famílias. A compreensão desses contextos foi fundamental para definir os Requisitos Prospectivos da Solução no Framework Semiótico. No total foram identificados 37 requisitos que foram apresentados em formato de histórias de usuários e organizados em seis níveis de formalidade. A representação no formato de histórias de usuário permitiu manter os aspectos humanos e sociais da solução, o vínculo entre as partes interessadas e os requisitos originados por elas, que, por sua vez, informaram os protótipos de média fidelidade. O requisito HU29 e HU26, por exemplo, deram origem a telas com funcionalidades de cadastro de várias fontes de canais e tipos de informação para permitir o acesso da população independente de sua condição. O requisito HU22, por sua vez, foi utilizado para a prototipação de um menu superior, com opções de acessibilidade.

Com este processo de design, fundamenta-se uma solução consciente e informada pelo contexto de saúde brasileiro, seus desafios e características. No design da solução, identificou-se a saúde como formada por redes de profissionais que trabalham de forma conjunta e a importância de relações e vínculos humanos entre profissionais e pessoas usuárias do SUS. Para agentes, o foco da solução está em fortalecer vínculos na comunicação em saúde e aproximá-los dos territórios. Para pessoas gestoras da APS, o foco está em apoiar o gerenciamento enquanto atividade fundamental e diretamente relacionada ao sucesso de uma campanha em cenários geográficos e socioeconômicos desafiadores.

As informações aqui sintetizadas não são finais e passarão por contínuo refinamento e evolução, a partir dos estudos realizados em trabalhos futuros e o aprofundamento da pesquisa. Esse estudo está inserido como um importante ponto de partida para um design que seja socialmente consciente e que crie intervenções tecnológicas que não impliquem em maior exclusão social e na negligência de valores humanos básicos.

Os próximos passos desta pesquisa incluem, não exaustivamente, as seguintes atividades:

i) estudos com partes interessadas do contexto para refinar o entendimento do contexto, com aplicação de técnicas como entrevistas, observação etnográfica e grupos focais;

ii) definição de diferentes perspectivas da solução em seus requisitos, modelos de software e arquitetura;

iii) workshops com partes interessadas do domínio para avaliar, levantar e especificar requisitos e funcionalidades para a solução, considerando principalmente as diferentes necessidades de partes interessadas em domínios socioeconômicos desafiadores;

iv) evolução da prototipação, materializando diferentes aspectos de uma extensão do e-SUS Território, assim como soluções tecnológicas multimodais no contexto da avaliação e feedback dos serviços de saúde pela pessoa usuária do SUS;

v) avaliação da solução com partes interessadas do domínio, colocando a proposta em operação em contexto situado real, entendendo e documentando aspectos de melhoria, realizando as respectivas sugestões de melhoria e refinamento.

Parte dessas atividades já está em andamento. Por exemplo, este estudo de diagnóstico foi relevante para definir objetivos e questões de pesquisa investigados em entrevistas semi-estruturadas com propósito exploratório. Duas entrevistas foram realizadas: com um ACS e um gestor da APS, que são partes interessadas centrais do domínio. Um projeto de pesquisa para apreciação do Comitê de Ética em Pesquisa da UFPR está sendo elaborado, definindo o planejamento das etapas futuras desta pesquisa com envolvimento de partes interessadas em atividades participativas e os respectivos instrumentos éticos.

Este estudo diagnóstico, os estudos empíricos e demais resultados a serem produzidos serão publicados em eventos e periódicos nacionais e internacionais, a fim de compartilhar os conhecimentos científicos, teóricos e práticos adquiridos na pesquisa. Os materiais também serão disponibilizados de forma aberta e acessível, na perspectiva de ciência aberta, para possibilitar amplo acesso e que outras pesquisas possam ser desenvolvidas no contexto investigado.

# Referências

# Apêndices

## Apêndice 1 - Artefato Quadro de Avaliação Preenchido

**Tabela.** Antecipação de Problemas para Partes Interessadas Identificadas no Contexto de Campanhas na Atenção Primária à Saúde e respectivas Ideias de Solução

| Stakeholder | Problemas e Questões | Ideias e Soluções |
|---|---|---|
| **Contribuição** | | |
| Pessoa Usuária do SUS | 1. Não ter acesso a meios de comunicação de massa.<br>2. Dificuldade de compreensão das instruções de campanhas de saúde, perdendo prazos, deixando de atender a chamados do SUS e/ou de tomar atitudes simples relacionadas a essas iniciativas.<br>3. Não ter dispositivos digitais como celulares e computadores pessoais, em um contexto em que serviços de saúde podem estar disponíveis apenas ou primariamente por meio de aplicativos.<br>4. Não conseguir fornecer feedback sobre os serviços do SUS devido à alguma condição individual.<br>5. Ter problemas de atendimento e não conseguir reportar:<br>5.1 Condições de atendimento que afetam a autonomia do paciente e/ou do responsável/familiar.<br>5.2 Médicos que não atendem no horário esperado.<br>5.3 Remédios entregues pela Farmácia Popular vencidos, próximos ao vencimento ou em falta.<br>5.4 Médicos pedindo pagamento por serviço gratuito.<br>5.5 Filas muito grandes para atendimento.<br>5.6 Demora no atendimento.<br>5.7 Os serviços de saúde primária não chegam à comunidade, em lugar de difícil acesso.<br>5.8 Falta de atendimento humanizado e educado para o paciente e a família ("preferem olhar para o papel e caneta do que para meu rosto").<br>5.9 Falta de condições apropriadas nos lugares de atendimento (unidades de saúde pouco acolhedoras, insalubres, sem estrutura de espera, acompanhamento ou descanso, ou falta de estrutura para o próprio atendimento de saúde).<br>5.10 Poucos médicos em relação à demanda.<br>5.11 Falta de leitos hospitalares e desigualdade na distribuição pelas regiões do País, bem como insuficiência de leitos de UTI.<br>5.12 Ausência de especialistas ou de serviços com exames e procedimentos nas proximidades do local de residência.<br>6. Desinformação sobre como funciona o serviço de saúde e como conseguir acessar o serviço.<br>7. Receber uma instrução incompleta, difícil de entender (ex.: com jargões técnicos excessivos), mal comunicada pelo profissional de cuidado da saúde.<br>8. Discriminação no atendimento.<br>9. Mesmo tendo um dispositivo que permite a instalação do Meu SUS Digital e outras aplicações do MS, o usuário tem acesso limitado à Internet e/ou intermitente (Ex.: não possui pacote de Internet no celular ou acesso à Internet em casa; a cobertura da rede da região de domicílio é de baixa qualidade, atrapalhando o uso de soluções que dependem da Internet). | 1-3. Facilitar o processo de comunicação das campanhas de saúde aos usuários do SUS, provendo diferentes recursos para transmissão das mensagens de forma inclusiva, acessível e com linguagem simples.<br>4. Oferecer soluções que contribuam com a autonomia da pessoa, ampliando sua capacidade de expressão, interação e exercício de seus direitos.<br>4. Contemplar diferentes modos de interação (ex: texto, voz, imagens, interfaces tangíveis, etc.).<br>5. Conceber soluções que fortaleçam a interação e o vínculo nas comunidades locais, não substituindo o contato substancial humano.<br>6. Comunicar segurança e confiabilidade do contato e das informações, de modo a conscientizar e evitar fraudes.<br>7. Contemplar diferentes níveis socioeconômicos, de experiência com dispositivos, de letramento, de habilidades.<br>9. Contemplar diferentes formas de comunicação (ex.: notificações, e-mail, SMS, ligação, chatbot).<br>9. Projetar soluções que funcionem no modo *offline* e que operem em dispositivos com baixa capacidade de processamento e armazenamento. |

| Perfil | Desafios | Oportunidades |
|---|---|---|
| Profissionais da Atenção Primária à Saúde (de forma geral) | 1. Dificuldade para operar os sistemas do SUS:<br>1.1 Problemas para manter registros de atendimentos atualizados e corretos.<br>1.2 Curva de aprendizagem complexa para aprender a usar os sistemas.<br>1.3 Dificuldades de uso e de acesso por diferentes questões (ex.: qualidade do acesso à Internet, habilidades de uso, etc.).<br>1.4 Equipamentos obsoletos.<br>2. Dificuldades com a infraestrutura de atendimento;<br>3. Receios de problemas com privacidade e segurança;<br>4. Falta de recursos de acessibilidade. | 1.3. Projetar soluções que funcionem no modo *offline*, que possam transmitir dados apenas quando conectadas em WiFi, e que operem em dispositivos com baixa capacidade de processamento e armazenamento.<br>1.4 Viabilizar soluções que possam ser utilizadas por profissionais via diferentes dispositivos.<br>4. Projetar soluções que sejam acessíveis e fáceis de usar, não onerando o tempo e esforço do usuário. |
| Agentes Comunitários de Saúde (ACSs) | 1. Conseguir atender uma população diversa e extensa (até 750 pessoas cobertas por um único ACS), com pouco contingente profissional.<br>2. Trabalhar sozinho, muitas vezes, visitando diretamente áreas que colocam o profissional em situação de vulnerabilidade, como regiões violentas, áreas rurais isoladas e distantes, com falta de infraestrutura e saneamento, em contato com doenças e epidemias.<br>3. Dificuldade em manter vínculos com famílias atendidas.<br>4. Danos psicológicos e emocionais por entrar em contato com sofrimento humano e injustiças vistas nas visitas às famílias.<br>5. Encontrar famílias que não querem se vacinar, não desejam acompanhamento de saúde, resistem a orientações ou não querem receber tratamento epidemiológico em suas casas.<br>6. Dificuldade para entrar em lugares fechados, como condomínios.<br>7. Falta de equipamento adequado para trabalhar com questões endêmicas (Máscaras, luvas e outros EPIs).<br>8. População não entende atribuição dos profissionais.<br>9. Falta de valorização da profissão.<br>10. Ter que percorrer longas distâncias sob sol, chuva, frio carregando materiais a pé ou utilizando recursos pessoais (ex.: bicicleta, moto).<br>11. Muitas mudanças domiciliares em suas microáreas, tendo que atualizar constantemente os dados demográficos de moradores.<br>12. Dúvidas sobre realizar o monitoramento dos pacientes de sua microárea: pode fazer tudo de memória, ao chegar na UBS? Pode usar anotações em papel? O que fazer quando buscar registrar algo via app e, por algum motivo, não conseguir concluir o registro? Como o ente público intervém para ajudar nessa tarefa de acompanhamento?<br>13. Dúvidas sobre como fazer o repasse dos seus acompanhamentos para outro profissional, caso mude de microárea. | 1. Auxílio no cálculo de rotas e estratégia de gestão de micro-áreas, com formação sobre atendimento humanizado e estratégias de comunicação.<br>2. Apoio da eMulti na atuação do ACS e acesso a Equipamentos de proteção individual (EPI).<br>3. Grupos digitais em redes sociais preferidas da população; criação de encontros presenciais para criar vínculos e momentos acolhedores entre equipe de APS e população local.<br>4. Apoio psicossocial aos ACS.<br>5. Campanhas de conscientização sobre a atuação dos ACS e temas centrais de desinformação, como a vacinação.<br>6. Mapeamento de centros que não recebem os ACS para articulação e diálogo.<br>7. Disponibilização e acesso a EPIs.<br>7. Central de alerta/denúncia ao MS-SUS sobre a falta de condições de trabalho.<br>8. Campanhas de conscientização sobre os próprios agentes de saúde.<br>10. Incluir recursos de deslocamento no planejamento de trabalho do ACS.<br>11. Funcionalidades que facilitem deletar e atualizar um domicílio nos sistemas de informação de saúde utilizados.<br>12 e 13. Definição de estratégias e compartilhamento de boas práticas entre a comunidade de ACS. |
| Agentes de Endemias | 1. População não entende atribuição dos profissionais.<br>2. Trabalhar, muitas vezes, sozinho visitando diretamente áreas que colocam o profissional em situação de vulnerabilidade, como regiões violentas, áreas rurais isoladas e distantes, com falta de infraestrutura e saneamento, em contato com doenças e epidemias.<br>3. Desvio de função e muitas atribuições.<br>4. Falta de valorização da profissão.<br>5. Ter que percorrer longas distâncias sob sol, chuva, frio carregando materiais a pé ou utilizando recursos pessoais (ex.: bicicleta).<br>6. Perda de materiais de trabalho devido às condições meteorológicas.<br>7. Dificuldade para entrar em lugares fechados, como condomínios.<br>8. Falta de equipamento adequado para trabalhar com questões endêmicas (Máscaras, luvas, material para coleta de larvas de aedes aegypti e outros EPIs).<br>10. Exposição e contaminação à doenças (ex: dengue, chikungunya, covid 19). | 1. Conscientização sobre o trabalho e papel dos ACE por meio de Campanhas e app Meu SUS Digital.<br>2. Apoio da eMulti na atuação do ACS e acesso a Equipamentos de proteção individual (EPI).<br>3. Fortalecimento da eMulti para atuação assertiva no trabalho e fortalecer canais de denúncia em caso de trabalho irregular.<br>5. Incluir recursos de deslocamento no planejamento de trabalho do ACS. |
| Gestores Municipais da Atenção Primária à Saúde | 1. Dificuldades no planejamento das ações da Atenção Primária à Saúde, bem como do gerenciamento e acompanhamento da atuação da equipe.<br>2. Dificuldades no manejo de recursos associados às campanhas.<br>3. Calendário de campanhas diverso que demandam ações imediatas, muitas vezes sem tempo de planejamento.<br>4. Dificuldade na manutenção de indicadores de produção (número de domicílios visitados, tipos de visitas sendo realizadas, vacinas aplicadas, etc.) associado às campanhas. | 1. Auxiliar no planejamento local da execução da Campanha, utilizando ou estabelecendo padrões, como o da vacinação de alta qualidade (BRASIL, 2023).<br>1. Solução que permite monitorar indicadores relacionados à execução da Campanha e do trabalho da equipe.<br>2. Descentralizar o manejo dos recursos para as ESF, definindo de forma situada o uso e prioridade do uso dos insumos dependendo do seu tipo (informacional, medicinal, etc.).<br>3. Seguir padrão de vacinação de alta qualidade e evitar atrasos na entrega dos insumos.<br>4. Planejar atuação no território, de forma a otimizar deslocamento, tempo e evitar retrabalho entre os agentes. |
| **Fonte** | | |
| Unidades Básicas de Saúde (UBSs) | 1. Falta de equipamentos e instabilidade de Internet.<br>2. Falta de informações sobre as normas/regulamentações do SUS.<br>3. Indisponibilidade de ofertas de serviços a todos os usuários que procuram a UBS.<br>4. Falta de profissionais nas equipes de atendimento.<br>5. Sobrecarga nas agendas.<br>6. Dificuldades de gerenciamento de equipes.<br>7. Alta rotatividade de funcionários.<br>8. Falta de medicamentos.<br>9. Sistemas não integrados à Rede Nacional de Dados em Saúde (RNDS).<br>10. Ausência de profissionais de suporte em TI. | 1. Subsidiar recursos para infraestrutura tecnológica das UBS.<br>2. Consolidação de informações em canal centralizado e confiável, por exemplo em um website facilmente acessível, uma seção do aplicativo e-SUS Território. Canal de dúvidas (FAQ) e perguntas do ACS/ACE com MS-SUS.<br>3. Fortalecimento de atendimento online para casos relacionados; gestão e otimização de agendamentos de consultas e atendimentos presenciais.<br>4. Fortalecimento do recurso de pessoal da saúde.<br>5. Otimização de calendários de Campanhas, definindo campanhas e suas respectivas agendas anuais para que UBS se preparem de antemão.<br>6. Treinamento e orientação em gerenciamento de equipes; compartilhamento de boas práticas e acolhimento de sugestões dos ACS, ACE e eMulti.<br>9. (Re)design de soluções adotarem padrões de interoperabilidade brasileira. |
| Farmácia Popular | 1. Sonegação fiscal e desvio de medicamentos.<br>2. Irregularidades nas vendas de medicamentos.<br>3. Problemas de estrutura física inadequada.<br>4. Ausência de farmacêuticos nas farmácias.<br>5. Controle de estoque deficitário.<br>6. Dificuldade de credenciamento de novas farmácias.<br>7. Sucateamento de material permanente.<br>8. Infraestrutura computacional inadequada. | 1. Fortalecimento de canais de denúncia que usuários do SUS podem utilizar.<br>2., 5. e 7. Definição de indicadores sobre ações fraudulentas e mal gerenciamento de recursos no SUS e monitoramento destes respectivos indicadores. |
| Meu SUS Digital | 1. Aplicativo com muitos usuários que usa autenticação via SouGov.<br>2. Centralização dos serviços em um único aplicativo possui vantagens práticas, porém pode dificultar o entendimento pelo usuário final e pode acarretar problemas de desempenho.<br>3. Desafios técnico de escalabilidade e interoperabilidade entre as aplicações vinculadas ao Meu SUS. | 1. Melhoria na usabilidade de autenticação por meio do SouGov, reduzindo passos e cliques necessários para autenticação (ex: uso de biometria, reconhecimento facial/video ou de voz).<br>2. Estudos de usabilidade, experiência de usuário e acessibilidade para refinamento e evolução das soluções existentes. Parceria com Sociedade Brasileira da Ccomputação e Comissões Especiais para realizar avaliações das soluções de saúde em larga escala.<br>3. Adoção de padrão de interoperabilidade em soluções novas ou |

|  |  | redesign de soluções existentes a partir do padrão supracitado. |
|---|---|---|
| **Mercado** | | |
| Secretarias Municipais de Saúde<br><br>Prefeituras | 1. Conseguir fazer acompanhamento adequado das pessoas em tratamento pelo SUS.<br>2. Necessidade de qualificação contínua das equipes gestoras municipais.<br>3. Necessidade de organização das portas de entrada do SUS.<br>4. Avaliação dos serviços prestados via SUS.<br>5. Necessidade de avaliação dos resultados das políticas municipais.<br>6. Garantia de prestação de serviços em seu território.<br>7. Prestação de contas de serviços ambulatoriais e hospitalares fornecidos na atenção primária à saúde.<br>8. Administração da oferta de procedimentos ambulatoriais de alto custo e alta complexidade.<br>9. Operação complexa dos sistemas de informações do SUS.<br>10. Avaliação do impacto das ações do SUS sobre as condições de vida e saúde.<br>11. Execução otimizada das ações de vigilância à saúde (sanitária, epidemiológica, zoonoses, controle de endemias, etc.).<br>12. Gestão dos servidores da saúde, próprios e municipalizados.<br>13. Dificuldades de administração financeira. | 1. Sistema de informação de saúde com acompanhamento fino da jornada e situação de saúde dos cidadãos.<br>4. e 9. Solução multimodal para receber feedback sobre atendimento no SUS, com diferentes modalidades de solução para atender diferentes condições físicas e socioeconômicas.<br>10. Estudos para proposição, avaliação e monitoramento de indicadores a respeito do impacto do SUS a partir do feedback adquirido na avaliação dos serviços pelo usuário do SUS. |
| **Comunidade** | | |
| Família, Acompanhantes | 1. Falta de privacidade e de individualidade.<br>2. Sofrimento psíquico sobre as dores, indefinição ou tratamento do paciente acompanhado.<br>3. Incerteza sobre o prognóstico/diagnóstico.<br>4. Ser tratado de forma não humanizada.<br>5. Dificuldades socioeconômicas para um acompanhamento efetivo.<br>6. Falta de acessibilidade geral no atendimento, acompanhamento e retorno ao serviço de saúde. | 1. Viabilizar que um mesmo dispositivo tecnológico seja utilizado para o acompanhamento de mais de uma pessoa.<br>1. Permitir o acesso e gerenciamento a aplicativos como MeuSUS por pais ou responsáveis de dependentes.<br>1. Garantir a privacidade e o tratamento adequado de dados pessoas e dados sensíveis.<br>2. e 4. Aplicação e monitoramento de protocolo de atendimento humanizado que reduz, pela empatia e acolhida, o sofrimento de pacientes e acompanhantes. |
| Ministério da Saúde, Estados e Municípios | 1. Desafios para a articulação entre os programas de saúde e a sociedade.<br>1.1 Concentração e falta de profissionais de saúde pelo território nacional.<br>1.2 Dificuldades em atender um país de dimensões continentais (8,5 milhões km²).<br>1.3 Grandes diferenças regionais (sociais, econômicas, culturais, epidemiológicas) que requerem atenção.<br>2. Dificuldades em garantir acesso à saúde qualificado para todas as pessoas.<br>3. Dificuldade na interação com os municípios para gerir e aplicar recursos públicos na saúde.<br>4. Complexidade na integração entre sistemas (técnicos e formais) municipais, estaduais e federais que podem dificultar o acompanhamento e oferecimento dos serviços.<br>5. Desafios em acompanhar avanços tecnológicos.<br>6. Desafios com o aprisionamento tecnológico, na manutenção da soberania tecnocientífica e necessidade de soluções que garantam a autonomia e a segurança no contato com as pessoas e na interação com seus dados. | 1. Projetar uma solução que possa se comunicar com diferentes sistemas nos níveis federal, estadual e municipal, respeitando padrões de interoperabilidade, software livre e gratuito.<br>5. e 6. Utilizar soluções baseadas em software livre.<br>5. e 6. Utilizar infraestrutura baseada em modelos, protocolos e tecnologias abertas, gratuitas e livres. |

Apêndice 2 - Relatório de Netnografia no Contexto da APS e Campanhas de Saúde

# Uma Netnografia Exploratória para Aprofundar o Entendimento sobre Desafios de Agentes e Campanhas de Saúde no Contexto da APS

Deógenes Silva Junior[1], Jonas Lopes Guerra[1], Krissia Menezes, Marisa Sel Franco[1], Roberto Pereira[1]

[1] Programa de Pós-Graduação em Informática – Universidade Federal do Paraná (UFPR)

Curitiba – PR – Brazil

{dpsjunior,jlguerra,kmlmenezes,msfranco,rpereira}@inf.ufpr.br

**Resumo**

*Este relatório apresenta os resultados de uma netnografia realizada em redes sociais públicas sobre o trabalho de Agentes de Saúde em sua atuação de promoção de saúde e em campanhas de saúde. Como metodologia, uma netnografia foi realizada, em que palavras-chave foram aplicadas em campos de busca em redes sociais, sendo os resultados coletados e analisados a fim de identificar tendências e temas que emergem na prática que esses profissionais compartilham publicamente online. Como principais resultados, tem-se quatro temas que ilustram a identificação de práticas de visitas domiciliares, iniciativas de promoção à saúde, a produção local de campanhas, e desafios relacionados à atuação, como a falta de materiais e recursos de trabalho, assim como a falta de reconhecimento.*

**Palavras-chave:** Agentes de Saúde, Campanhas de Saúde, Atenção Primária à Saúde, Netnografia.

## 1. Contexto

O projeto "Projeto Ministério da Saúde - Comunicação com os usuários do SUS", com Termo de Execução Descentralizada número 173/2023 e 175/2023, tem como objetivo "*a comunicação direta entre usuários e a Atenção Primária à Saúde (APS) por meio da pesquisa, desenvolvimento e inovação em tecnologias de comunicação segura e acessível, armazenamento de dados e monitoramento das relações com os usuários do Sistema único de Saúde (SUS)*". No contexto deste projeto, o Design Socialmente Consciente (Baranauskas et al., 2024) foi adotado como a abordagem teórico-metodológica desta pesquisa.

De acordo com o Design Socialmente Consciente, toda construção de tecnologia deve considerar de forma ampla as partes interessadas do domínio de forma participativa e universal, entendendo valores, necessidades e expectativas das partes interessadas antes de se produzir qualquer solução (Baranauskas et al., 2024). Com isso, o design de uma tecnologia é realizado de forma consciente do contexto social e das pessoas que o constituem. A literatura (Ferrari et al., 2020; da Silva Junior et al., 2024) recomenda realizar estudos de empatia e conhecimento de contexto antes de envolver as partes interessadas diretamente em atividades participativas. Esse cuidado é importante como forma de responsabilidade ética e cuidado com as partes interessadas, ainda mais em contextos sensíveis, como é o caso da saúde no Brasil.

A etnografia é uma das maneiras de se realizar estudos sobre um contexto social, se sensibilizando sobre as práticas das pessoas, suas opiniões e formas de expressão. "Projetos de pesquisa etnográfica usam imersão profunda e participação em um contexto de pesquisa específico para desenvolver uma compreensão que não seria alcançável com outras abordagens de pesquisa mais limitadas" (Lazar et al., 2017, p. 230). A Atenção Primária à Saúde no Brasil ocorre em diversos contextos socioeconômicos, onde cada região geográfica pode possuir condições específicas de trabalho. A netnografia, como uma forma de etnografia que permite entender como as comunidades se organizam em grupos sociais online e quais informações compartilham (Kozinets, 2002), surge como uma opção que permite pesquisadores entenderem e se sensibilizarem sobre práticas oriundas de contextos geograficamente dispersos.

Para exemplificar, foi possível identificar na netnografia a atuação social e política de Agentes de Saúde para luta por direitos, que são uma categoria profissional especialmente engajada em sindicatos (Batistella, 2013). A partir de sua mobilização presencial e online, Agentes de Saúde obtiveram conquistas recentes, como a aprovação do relatório do PL 2113/2022 que torna obrigatório o pagamento do piso salarial dos Agentes de Saúde, e a Sanção Presidencial do PL 2012/2019, que prevê a indenização de transporte ao Agente Comunitário de Saúde (ACS) e ao Agente de Combate às Endemias (ACE) como forma de custeio de locomoção.

Deste modo, este relatório descreve uma pesquisa netnográfica em redes sociais, com o objetivo de promover um processo de sensibilização sobe práticas, expectativas e desafios amplos de agentes de saúde no contexto da APS do Brasil. Foi definida como redes social o Facebook, com publicações veiculadas publicamente pelos próprios agentes de saúde na internet, já que o Facebook possibilita criar fóruns públicos de organização social. Os dados coletados (publicações e comentários) foram analisados de forma iterativa e incremental em abordagem qualitativa, identificando temas que surgiram qualitativamente nas postagens e aspectos da realidade que eram retratados pelos agentes em suas postagens.

Foi possível identificar, com a netnografia, três temas que ilustram práticas de promoção de saúde de Agentes por meio de visitas domiciliares individuais e em conjunto com uma equipe de saude da família (eSF); iniciativas locais de criação de materiais, folders e panfletos relacionados à Campanhas de Saúde; a veiculação de palestras que ocorrem no contexto das UBS e de escolas; desafios relacionados à falta de infraestrutura, locomoção, falta de materiais de trabalho, e à luta por reconhecimento profissional.

Este relatório está organizado da seguinte forma: a Seção 2 apresenta a metodologia desta pesquisa netnográfica. A Seção 3 apresenta os resultados, caracterizando o corpo de dados coletado. A Seção 4 apresenta a discussão dos temas e tendências encontrados nos dados. Por fim, a Seção 5 apresenta as considerações finais.

## 2. Metodologia

Comunidades *online* têm uma existência real para seus participantes e podem afetar o comportamento das pessoas em atividades *online* e *offline* (Ivan, 2019). A etnografia é "a abordagem da etnografia aplicada ao estudo de culturas e comunidades online" (Kozinets, 2010) em que o envolvimento pessoal com a comunidade é a forma de revelar as perspetivas quotidianas dos seus membros e de compreender a comunidade "a partir de dentro" (Ivan, 2019).

A abordagem netnográfica permite-nos observar: com quem os participantes do grupo social comunicam usando ferramentas online e também sobre quais tópicos e sobre que tipos de informação comunicam (Ivan, 2019).

Nós baseamos nossa pesquisa nos três passos de netnografia recomendados de Ivan (2019): (1) Entrada — o estabelecimento de questões de pesquisa e identificação e seleção da comunidade online, incluindo obter acesso à comunidade; (2) Coleta de dados — observação de membros da comunidade, arquivamento e revelação das interações sociais (consocialidade), com preocupação em conduzir pesquisas éticas; (3) Análise e interpretação — classificação, codificação, contextualização e significados, incluindo oportunidades de feedback de membros da comunidade.

Esta netnografia pode ser classificada como observacional, quando não houve praticamente nenhum envolvimento com a comunidade online (Ivan, 2019). A seguir, apresentamos a estratégia de investigação a partir dos três pontos.

*Entrada.* O objetivo da pesquisa é entender as práticas e desafios profissionais de Agentes de Saúde. Para isso, utilizamos os seguintes critérios de seleção dos fóruns para a netnografia: redes sociais onde Agentes de Saúde pudessem se reunir em comunidades virtuais e publicar opiniões, impressões, dúvidas sobre sua profissão. Dentre as redes sociais mais utilizadas no Brasil, o Facebook foi escolhido por sua funcionalidade de criar grupos e permitir agrupamentos de pessoas em torno de assuntos específicos. Outras redes sociais, mais voltadas para veiculação de publicações sem funcionalidades de organização social não foram consideradas (ex., Instagram e TikTok). No Facebook, caso a publicação realizada tivesse *links*, o *link* seria visitado e registrado, quando apresentasse informações relevantes para a atuação de agentes na APS. Não foram coletados dados pessoais ou quaisquer informações fora do escopo da atuação na APS.

*Coleta de dados.* A coleta de dados incluiu a busca por palavras-chave nas funcionalidades de busca do Facebook, navegando por todos os resultados e registrando em um documento capturas de telas dos publicações que tratavam sobre o tema de interesse. Quando a publicação tinha comentários de outros usuários, os comentários eram lidos e também registrados, caso relevantes. Como *strings* de busca, utilizamos palavras-chave abrangente para retornar a maior quantidade possível de resultados. Foram utilizadas as seguintes strings de busca: "agente comunitário de saúde", "agente de combate à endemias", "campanhas de saúde", "dificuldades ACS", "dificuldades ACE". Todas as postagens retornadas pela rede social foram analisadas.

*Análise e interpretação.* Os dados foram analisados, primeiro transcrevendo as postagens textuais e transcrevendo os elementos visuais que aparecem em figuras, infográficos e banners presentes nas publicações de redes sociais. A transcrição de texto foi realizada de forma anônima, não revelando localizações, nomes ou títulos. A partir do texto, foram gerandos códigos. Os códigos interativamente e progressivamente sendo revistos e refinados à luz da análise dos dados. A partir do agrupamento dos dados, foram criadas categorias que representavam temas que surgiram nos dados.

Esta pesquisa não busca trazer conclusões finais a partir da netnografia, mas ela deve ser entendida como um dos métodos utilizados em uma investigação mais ampla. A netnografia permitirá, a partir da combinação com outros estratégias de investigação, possuir um melhor entendimento sobre a atuação de agentes de saúde no contexto diverso da APS no Brasil.

## 3. Resultados

A análise realizada formou um corpo de sessenta e cinco (65) publicações capturadas no Facebook e cinco capturas de tela de um blog de um agente de endemias (encontrado durante as publicações do Facebook). Dentre o corpo de informação desta análise, 11 capturas de tela eram referentes a troca de comentários entre usuários na plataforma. Um total de 169 códigos foram gerados na análise.

Caracterizando a fonte das publicações, elas foram identificadas em páginas de Secretarias de Saúde, de Prefeituras Municipais e de páginas pessoais de Agentes de Saúde. Foi possível identificar publicações de diferentes regiões do país, sendo identificado secretarias e municípios do doze (12) estados: Acre, Pará, Rio Grande do Sul, Rio de Janeiro, Espírito Santo, Paraíba, Amazonas, São Paulo, Paraná, Santa Catarina, Goiás e Bahia. Alcançar a diversidade de estados era um objetivo desta netnografia, visto que as comunidades online se manifestam a partir da realidade de pessoas de diferentes regiões do país. Com isso, atinge-se uma diversidade de condições e situações de saúde e de atuação em saúde, importante para o objetivo exploratório deste estudo.

Foram identificadas publicações caracterizando situações ribeirinhas, de comunidades de zona rural (como fazendas), de espaços urbanos, de capitais e de municípios do interior do estado. Apareceram espaços físicos nos quais as partes interessadas estavam em UBS, em centros de saúde, escolas, condomínios, domicílios, quintais, ruas pavimentadas e não pavimentadas e canoas em rios. Essas condições evidenciam que uma busca exploratória em uma única rede social já foi capaz de evidenciar a diversidade de condições de saúde identificadas no contexto brasileiro. Cada condição de saúde oferece cuidados e desafios únicos em relação à APS e às campanhas.

O corpo de informações se refere principalmente de realidades de agentes de saúde, como ACS e ACE, mas na maior parte das publicações acompanhados de uma equipe de saúde multidisciplinar: técnicos, enfermeiros, médicos, nutricionistas, dentre outras especialidades médicas. Partes interessadas no contexto da gerência da APS e das campanhas também apareceram nas publicações, como municípios e secretarias municipais. Por fim, usuários de saúde do SUS apareceram em fotos, por exemplo dentro de UBSs. Deste modo, a análise de dados evidencia a diversidade de partes interessadas afetadas e que influenciam o contexto da APS e das campanhas de saúde. Diversas partes interessadas formam uma rede de relações de parceria profissional (entre profissionais de saúde), de cuidado (entre os próprios profissionais, mas principalmente com a população de microáreas e territórios) e de relações sociais amplas.

As campanhas que apareceram no conjunto de dados abrangem campanhas relacionadas a meses temáticos (ex., outubro rosa e novembro azul), diversas campanhas de vacinação, campanhas relacionadas a ações voltadas à prevenção do HIV e outras ISTs, e campanhas relacionadas a covid-19, dengue, malária e outras arboviroses. Também foram identificadas campanhas relacionadas a abuso infantil.

Apresentamos os temas que emergiram no conjunto de dados analisados. Os temas são específicos ao conjunto de dados estudado, limitando-se a essas realidades. É interessante para a análise aparecer e emergir temas e situações que representam casos únicos da realidade brasileira do que uma dimensão quantitativa de um determinado fenômeno aparecer de forma extensiva nos dados. A apresentação dos temas a seguir não buscam generalizar a realidade de contextos situados específicos. Antes, servem como exemplificação de indícios da variedade de condições que podem ser enfrentadas, que nos permitem, como trabalho exploratório, investigar em novos estudos, como investigações empíricas com partes interessadas e em relações com evidências da literatura.

## 4. Apresentação e Discussão dos Temas

A análise revelou três temas principais nos dados, que refletem características da APS e das campanhas de saúde, a maneira e o conteúdo da apresentação de agentes online, assim como desafios que agentes precisam enfrentar em sua prática. Os temas sao explicados a seguir, juntamente com códigos que exemplificam o tema.

### 4.1 Caracterização da Atuação na APS e nas Campanhas

Este tema representa as diversas maneiras como a APS se organiza no encontro com a população, no planejamento e execução de campanhas, assim como outras ações amplas de saúde realizadas no território por agentes de saúde e demais membros das equipes associadas, como ESF.

**Meses Temáticos de Campanhas.** Verificou-se a relação das campanhas de saúde com os meses do calendário. Aproveitam-se a combinação de "meses coloridos" para veicular mensagens específicas. São diversos eventos anuais associados com cores, como: janeiro branco (saúde mental); fevereiro roxo (alzheimer, fibromialgia e endometriose); Marco Azul-marinho (câncer colorretal); Abril Azul (Autismo); Maio Vermelho (Câncer de boca); Junho Laranja (Anemia e leucemia); Agosto Verde-claro (Linfoma); Setembro Amarelo (Prevenção ao suicídio). Para os últimos três meses do ano, foram encontradas evidências na pesquisa:
- Outubro Rosa (Câncer de mama): *"Nesta sexta feira (...) tivemos nosso evento do Outubro Rosa, que é o mês de conscientização da saúde da mulher e da prevenção ao câncer de mama."*;
- Novembro Azul (Câncer de próstata): *"fazendo parte da Campanha de Novembro Azul, que dedica a oferecer serviços e orientações para a saúde do homem, a prefeitura XXX através da Secretaria Municipal de Saúde está desenvolvendo diversas atividades para serviços de saúde do homem;*
- Dezembro Laranja e Vermelho (Câncer de pele e IST)[7]: *"A Secretaria de Saúde de XXX está promovendo a Campanha Dezembro Vermelho e, para isso, oferecerá horários estendidos no SAE (Serviço de Atenção Especializada) e nas UBS."*

Também identificou-se campanhas locais, iniciadas pela própria APS local. Por exemplo, uma campanha foi realizada com orientações de saúde em relação a dengue voltadas unicamente para gestantes (grupo vulnerável). *"A Prefeitura (...) deu início à campanha "Proteção Para Quem Está Gerando Vida", uma iniciativa inovadora que busca proteger gestantes, grupo considerado vulnerável, contra os riscos da dengue".* Uma suposição é que o município produziu uma campanha específica motivado por um número expressivo de gestantes localmente, ou a situação de que os lugares de moradia dessas gestantes estavam sendo afetadas por casos de dengue.

Ação realizada na prática: colar cartazes em pontos diferentes da cidade, como postes de passagem de pessoas, e lugares de comércio. Os panfletos relacionados às campanhas de saúde foram analisados. Foi possível perceber informações que aparecem no poster criado pela equipe de APS:
- Nome da equipe de ESF,
- Nome ou título da ação (ex: coleta de preventivo);
- Mensagem de conscientização (A prevenção é o melhor remédio)
- Dia(s) e horário(s) da ação.
- Público-alvo (mulheres de X a Y anos).
- Serviços oferecidos (ex: Exame de Sangue - PSA)
- Local (ex., UBS).
- Fotos das pessoas da ESF.
- Uso de padrão visual (paleta de cores, tipografia, ícones e figuras).

---

[7] https://www.correiobraziliense.com.br/revista-do-correio/2024/01/6774450-meses-coloridos-campanhas-de-prevencao-de-doencas-preenchem-o-calendario.html

Os panfletos colocados em postes e paredes eram no formato de papel A4. No sentido dessa ação de levar a informação até a população, os agentes chamam a si mesmos de "agente multiplicadores".

**Mutirões e Dia D de divulgação.** Muitas das ações de campanha ocorrem por meio de mutirões ou um chamado "dia D", na qual a APS organiza um *"dia D na microárea"*, geralmente realizada dentro das UBS, centros de saúde ou em infraestruturas montadas nas imediações das UBS (ex., tendas). Foi identificado que muitas vezes, o dia D é realizado ao final de campanhas, intensificando ações na reta final (*"marcou o encerramento das Campanhas Novembro Azul e Novembro Vermelho"*).

O dia D concentra e intensifica diversas ações de saúde que ocorrem em um local específico, realizando uma diversidade de serviços de saúde, além da conscientização e orientação por meio de palestras: *"As campanhas contaram com consultas, exames e avaliações oftalmológicas e odontológicas, com ênfase ao combate de câncer de próstata e bucal (...)"*. O rol de serviços realizados é extenso, relacionados diretamente com o público-alvo e a situação de saúde relacionada com uma campanha. Nessas situações, profissionais orientam a população em relação às suas especialidades: *"O evento contou com a presença do Dr. (...), que orientou a população masculina, respondeu a dúvidas e reforçou a importância da prevenção"*.

Foram identificados serviços de saúde, odontológicos e de orientação que são ofertados em dias associados à campanha ou em um único dia (o dia D), como *"testes rápidos, medição antropométrica, solicitação de exames, avaliação da carteira de vacinação, verificação da pressão arterial e glicemia, testes rápidos de HIV, Sífilis e Hepatite B e C, medidas antropométricas (peso e altura), circunferência da cintura e do quadril, exame PSA, avaliação da carteira de vacinação."* A campanha de Novembro Azul, por exemplo, realizou *"cerca de 100 exames de PSA (...), além de palestras, orientações e consultas com médicos especialistas."*

Mesmo as ações de mutirão também podem se espalhar por diversas localizações de um mesmo território, indo de uma UBS para outra: "a *campanha segue com novas atividades programadas: Multirão no (...), no dia (...), das Xh às Yh, a Unidade de Saúde do bairro oferecerá exames de PSA e testes rápidoss para sífilis e HIV. (...) Dr. XXX: (...) ministrará uma palestra voltada à conscientização e prevenção do câncer de próstata. Multirão no (...) a Unidade de Saúde do bairro estará aberta para atender a população masculina com exames de PSA e testes rápidos."* Com isso, capilariza-se o atendimento à população, promovendo ações especificas para cada localidade.

As equipes de saúde saem no território, realizando ações de divulgação em locais estratégicos, por meio de falas, divulgação e colagem de panfletos: *"Hoje foi dia de andar* *por todo território de nossa equipe para divulgarmos o dia especial da coleta de preventivo para as mulheres cadastradas em nossa equipe".*

Nestes momentos de multirão ou dia D, quando ocorrem no ambiente da UBS, envolvem a preparação do local com cuidado para receber a população. Foram utilizados materiais temáticos relacionados a cada campanha, como decoração, balões coloridos, cartazes, *banners*, e até camisetas temáticas da campanha usadas pelos profissionais de saúde (ex., camisetas com identidade visual sobre o outubro rosa).

**Visão Ampla de Saúde.** Foi possível identificar uma atuação da APS e de agentes não voltada para uma saúde voltada para doenças, mas uma visão ampla que inclui uma integralidade do cuidado, em promoção de bem-estar e boas práticas de vida. Nas campanhas realizadas no mês de outubro (câncer de mama) e novembro (câncer de próstata), por exemplo, não focam apenas no câncer e na doença, mas buscam promover hábitos saudáveis de vida (como atividade física), palestra com psicóloga (auto cuidado), assim como a promoção de ações socioafetivas de lazer e entretenimento (sessão de fotos, música ao vivo, sorteio de brindes, lanches da tarde e cafés da manhã): *"oferecemos palestras com uma psicóloga convidada sobre a importância do auto cuidado, com a enfermeira sobre a prevenção do câncer, atividade física com educador físico, e para descontrair sessão de fotos, música ao vivo, sorteio de brindes e um delicioso café da manhã completo!!"*

Um outro exemplo ilustra uma ação de conscientização e orientação à comunidade em relação a problemas não relacionados com doenças, mas com problemas socioambientais, como a violência infantil: *"(...) duas importantes ações de conscientização sobre o abuso infantil foram realizadas nas unidades de saúde XXX e YYY, com o objetivo de orientar a comunidade em geral sobre como prevenir e agir diante de casos de abuso sexual em crianças e adolescentes".*

A visão ampla de saúde também está associada com o compartilhamento de momentos positivos e de alegria entre as partes interessadas. Momentos de descontração são uma forma de promoção de maior integração da equipe de APS com a comunidade. Foi possível identificar, por exemplo, da comemoração de aniversário de uma ACS junto de um evento de campanhas com a população e outros profissionais de saúde: *"no final ainda tivemos um parabéns surpresa dedicado a uma de nossas Agentes Comunitárias".*

A visão ampla de saúde inclui a vigilância à saúde, que envolve dimensões socioambientais, como o controle vetorial e ações de combate às arboviroses, mesmo em casos em que não há presença direta de seres humanos: *"trabalho com a leishmaniose e tenho que colocar armadilha [para vetores, como roedores] todos os dias (...) no meu município são raros os casos humanos, mas tem uma quantidade considerável de animais positivos."* Em muitos casos, mesmo momentos voltados para campanhas de saúde (como as de combate ao câncer) eram oportunidades para a veiculação de mensagens relacionadas à vigilância à saúde e sanitária: *"ao final da programação o médico Doutor (...) fez um alerta com orientações ao combate da dengue, lembrando da responsabilidade das pessoas com a limpeza de seus quintais".*

**Atuação multiprofissional.** Foi possível identificar a atuação multiprofissional e realizada por equipes, não realizada somente por agentes de saúde ou outros profissionais de forma isolada. Foi possível observar como diferentes equipes percorrem o território *"Hoje foi [o dia] de andar por todo território de nossa equipe".*

Embora o território seja conhecido e relacionado diretamente com o trabalho de agentes, há momentos como o atendimento de acamados em que é necessário o engajamento ativo de outros profissionais indo diretamente para o território: *"Durante a* *semana que passou, a equipe XXXX concluiu o roteiro de vacinação da Influenza nos pacientes acamados ou com dificuldade de se dirigir até a UBS e que fazem parte da sua área"*. Nestas visitas, os profissionais aproveitam a presença na casa das pessoas para realizar uma ação intensiva de saúde: *" (...) diversas visitas domiciliares onde além da aplicação da vacina, foi aferido a pressão de todos os pacientes bem como muitas outras orientações sobre saúde"*.

Equipe de ACE, por sua vez, realizam ações que envolvem uma chamada "equipe de endemias": *"Equipe de Endemias no combate à malária realizando vigilância em saúde, busca ativa e coleta de lâminas na região XXX, devido o surgimento de casos positivos."* Também foi identificada um "equipe de vacinação antirrábica", realizando vacinação de animais: *"Mais um dia ao lado da equipe de vacinação antirrábica!"*. Na atuação de vigilância, muitas vezes os agentes aproveitam para realizar ações de avaliação e orientação relacionadas a mais de uma arbovirose e zoonoses: *"me contem, quem aí também se vira nos 30 e trabalha mais de um programa de prevenção quando é escalado?"*. Em um exemplo identificado, por exemplo, um ACE realizou em uma única visita a um domicílio a checagem da presença de morcegos, do fechamento do tanque de água e verificação de larvas do mosquito da dengue.

Parceria com escolas também foram realizadas, onde equipes formadas por enfermeiros, técnicos, médicos e agentes de saúde iam até o ambiente da escola realizarem ações de promoção à saúde. Em um caso específico identificado, uma ação da equipe de saúde em parceria com a escola foi realizada para trabalhar *"o maior problema ambiental detectado na comunidade: lixos domésticos."* Diversas ações foram realizadas no âmbito da própria comunidade local, na qual estudantes e professores atuaram como protagonistas e difusores de informação e ações concretas de saúde: *"durante toda a semana alunos puderam se aprofundar ao problemas com varias atividades: Palestra, Apresentação, Recolhimento dos lixos, Criação de lixeira e Distribuição de folder aos moradores para uma reflexão sobre o consumo excessivo, a produção de lixo e seu descarte correto"*.

Outro momento onde essa equipe multiprofissional sai em direção ao território ocorre nas visitas relacionadas a localizações geográficas remotas. Foi identificado, por exemplo, uma visita domiciliar em uma comunidade ribeirinha, por uma equipe de saúde da família composta de enfermeira, técnico de enfermagem, ACS e motorista. Na visita, oferecem serviços, acompanhamentos e ações de conscientização de saúde diversas: *"Aconteceu mais uma visita domiciliar na Comunidade XXXXX pela equipe YYY composta pela enf[ermeira] AA, Tec. de Enf. BBB, ACS X e Motorista Y. Foram ofertadas consultas de enfermagem para usuários, (...) avaliação de puericultura (...) dispersão de medicação, Solicitação e Avaliação de Exames, Coleta de Preventivo, Limpeza O[donto]lógica, Testagem Rápida e busca ativa de pacientes com comorbidades e dificuldades de se locomover"*.

## 4.2 Apresentação dos Agentes de Saúde nas Redes

**Utilizar a própria rede como forma de conscientização e divulgação de ações**
A própria rede social é utilizada por agentes de saúde como forma e canal de comunicação com a população, a partir da criação de publicações orientativas ou informativas. *"Tosse por 3 semanas ou mais Febre vespertina Sudorese noturna Emagrecimento Procure sua equipe de saúde da família Tuberculose tem cura!"* Informações importantes são veiculadas nas redes sociais dos agentes, utilizando linguagens acessíveis à população, com mensagens diretas e assertivas. Na perspectiva de ACS, são veiculadas informações sobre condições de saúde que ajudam a população a entender possíveis sintomas e situações que seu círculo social pode estar passando: "*O câncer de próstata, conhecido por ser silencioso nos estágios iniciais, pode apresentar sinais como dificuldade para urinar, necessidade frequente de urinar à noite, dor na região pélvica ou nas costas, e em casos mais avançados, dor óssea*". Na perspectiva de ACE, por sua vez, orientações sobre possíveis vetores de doenças são informadas à população: "*Ao encontrar morcegos mortos na sua casa entrem em contato com a unidade de vigilância de zoonoses da sua cidade imediatamente*"

A rede social também é utilizada como forma de divulgação e chamamento de ações realizadas, como multirões e "dias D". Foram identificados vários convites e chamamentos para participar das ações e vir presencialmente na UBS: "*Se você é morador da XXXX não deixe de comparecer à nossa unidade de saúde para consultas, exames, vacinas e eventos como esse!*" Informações sobre como acessar o serviço e o que é necessário levar para a UBS são veiculadas: "*Para ser atendido, é necessário apresentar um documento de identificação com foto.*" ou "*Para participar das ações, é necessário apresentar o cartão do SUS e um documento de identidade.*" Deste modo, as redes sociais se colocam como um importante canal que pode ser aproveitado para veicular mensagens de saúde e cuidados amplos: "*Homem que se cuida, se ama! A Secretaria de Saúde reforça a importância de procurar a unidade mais próxima para avaliações regulares, porque cuidar da saúde é um compromisso para o ano inteiro*".

**Utilização de memes e estratégias de redes sociais que comunicam sobre a prática e realidade profissional.** Foi verificado como agentes de saúde comentam sobre sua situação online com bom-humor, criando memes em imagens e vídeos que brincam com desafios e padrões de sua atuação junto à população, no trabalho na APS e em aspectos mais amplos sobre a profissão. Os memes são uma fonte de compartilhamento e engajamento entre os próprios agentes. Percebe-se a adoção de ferramentas e do tipo de veiculação informacional característico da plataforma também sendo adotados por agentes de saúde para comentar sobre sua realidade. Esta característica indica que esses profissionais poderiam criar o mesmo tipo de informação para, ao invés de apenas compartilhar aspectos de sua profissão, informar, orientar e conscientizar a população de seu território.

**Apresentação dos próprios ACS sobre (esclarecendo) sua profissão.** A literatura indica que muitas vezes a população não entende o trabalho de agentes de saúde (Evangelista et al., 2018). Nas redes, identificamos que muitas mensagens eram acompanhadas de explicação sobre a atuação e profissão de agente de saúde (ACS ou ACE): "*O ACS (...) [é] responsavel por gerir essas idas aos pacientes que possuem dificuldade de locomoção ou são acamados*". No caso da vigilância em saúde, o trecho a seguir ilustra a explicação sobre importância da atuação de ACE: "*A vigilância é imprescindível para evitar um surto ou incidência no aumento de casos.*"

A explicação da profissão era acompanhada da importância que essa atuação tinha na saúde das comunidades, famílias e espaços sociais: "*O ACS é peça importantíssima na ESF, capaz de levar a informação até a casa do paciente (...) [t]ornando-se um agente multiplicador*". Em outra publicação, um ACS indica que eles são "*(...) peça fundamental nessa ligação, pois ele tem o controle da área e sabe da necessidade do território orientando os demais profissionais da equipe ao lugar de atendimento*".

As publicações também era acompanhadas de uma expressão de orgulho e importância que os próprios agentes sentem em relação à profissão no contato humano direto com as famílias. Uma importância é indicada principalmente na atuação junto das pessoas com pessoas em situação de vulnerabilidade, como pessoas com dificuldade de locomoção e acamados (que não conseguem ir para a UBS) e levar serviços para lugares longínquos, como domicílios de zonas rurais: *"Caminhos que muitas vezes somos os primeiros ou os únicos a levarem os serviços públicos".* Além disso, mensagens e informativos comemorativos sobre o próprio dia nacional dos ACS e ACE (04 de outubro) também eram veiculados e compartilhados pelos agentes.

Deste modo, identifica-se uma prática existente dos profissionais explicarem para a população suas atribuições, responsabilidades e ações sendo realizadas, assim como reforçarem a importância de sua profissão. Esse comportamento pode ser aproveitado por uma solução em APS, na qual ajuda a população a compreender a atuação e importância dos agentes de saúde.

**Acolhida entre os Agentes e mobilizações pela luta por direitos.** Agentes de Saúde se articulam e formam comunidades online, se conhecendo, seguindo e comentando sobre aspectos relevantes em sua atuação. Um dos temas que emergiu foi referente ao curso técnico oferecido aos agentes de saúde no Brasil "Programa Mais Saúde com a Agente": *"Aces e Acs me contem: Já concluíram o seu curto Técnico? Me digam se vcs tbm [possuem] alguma graduação e o nome da cidade de vcs!".*

No contexto de condições adversas enfrentadas pelos agentes durante a atuação nos territórios, em interações sociais na plataforma os próprios agentes se expressavam sua identificação com o trabalho e se acolhiam: *"E uma luta mas n[ó]s gosta[mos] disso… de ser o que somos. (...) não podemos deixar a imunidade baixar e [a]quele chá da v[ó]vo com alho, limão e toda as folhas que ela conhece nos mantém de pé."*

Há a retomada de ganhos históricos que agentes de saúde conseguiram em sua luta histórica pelo reconhecimento da profissão: **"***Muitos não sabem, mas nós Agentes de Endemias e Comunitários de Saúde fomos elevados ao longo dos anos ao nível Técnico. Um grande avanço para a categoria."*

Também há uma troca entre agentes online principalmente na resolução de dúvidas sobre a profissão, sobre a diferença entre realidades situadas de agentes e na troca de experiências. Muitas dúvidas de agentes de saúde são informadas e negociadas online com outros agentes, principalmente sobre direitos trabalhistas e os limites de sua atuação (quais são minhas atribuições e quando é um desvio de função).

Para ilustrar, um diálogo foi identificado entre um ACE no contexto de manifestação de dúvidas, reclamações e condições de trabalho não propícias sobre o transporte de agentes em localidades longínquas: *"Por lei temos direito a transporte público. Se o município estiver irrregular com relação a isso Vcs n são obrigados a pagar passagem. Para grandes deslocamentos ou o município dá o transporte ou vcs n vão."* Neste ponto, o agente de saúde pergunta ao outro: *"Vcs tem sindicato?"*. O papel de sindicatos, como o CONACS, é central na mobilização de agentes de saúde pela luta por direitos e pela não retirada de direitos já conquistados (como o próprio transporte). Muitos agentes realiam chamentos para outros profissionais da categoria somarem à causa: *"temos que lutar pela desprecarização e pelo plano de cargos e carreiras pago pelo governo federal".*

**Valores relacionais expressos pelos próprios agentes.** Os agentes de saúde expressaram diversos valores e aspectos de importância que eles próprios comunicam para sua profissão e atuação. Postagens eram realizadas expressando o orgulho e valor que as pessoas tinham por trabalhar como agente de saúde: *"Deus disse: você nasceu para servir* *ao próximo e faço isso com muito amor, dedicação e orgulho (...) orgulho de poder ajudar ao próximo e saber que as mais de 150 famílias da zona rural podem sempre contar comigo, a qualquer dia, a qualquer hora."* Outras postagens e comunicações (por meio de comentários) eram acompanhadas de expressões espontâneas de valor: *"É gratificante (...) Mas no geral é um[a] benção poder ajudar a população"*, que expressam uma seus valores e afeto em relação ao trabalho: *"O que podemos ajuda[r] nos ajudamos com gosto"*.

Como valores identificados, percebeu-se o valor de disponibilidade, em que os agentes que se colocam disponíveis e acolhem as famílias: *"Estou sempre ali para tentar resolver seus problemas até que estejam aí meu alcance"*. Um valor central é o de acolhimento, do ACS com as famílias e vice-versa: *"É gratificante saber que em 22 anos de ACS as portas das casas de cada uma família sempre estiveram abertas para mim."* O valor de acolhimento aparece sendo expressado com vários sinônimos, como empatia, segurança e conforto: *"colocando-se no lugar do outro (...) Ser ACS é passar segurança para aquela família, de que o problema dela será sanado em breve, junto com a equipe da ESF… é passar conforto e acolhimento"*.

Esses valores também podem ser entendidos como expressões de habilidades importantes para o profissional. Um ACS, por exemplo, comenta sobre a resolutividade necessária na visitas domiciliares e aplicações de campanhas. Não basta apenas levar a informação e realizar (atualização) de cadastros. Uma resolutividade de ACSs são demandadas, indicando que esse profissional pode ser cercado diariamente com uma tomada de decisão que tem que ser realizada considerando diversos fatores de uma família: *"Ser Agente Comunitário de Saúde não é só sobre visita domiciliar… Ser ACS é saber lidar com as adversidades pré existentes (...) na nossa microárea, é buscar sempre uma resolutividade para uma ocorrência de Saúde, e sempre colocando-se no lugar do outro"*.

Por fim, esses valores como um todo se colocam em um contexto relacional, dos agentes e com a população, mas também entre agentes e outros profissionais de saúde. Os agentes expressam o valor de conseguir viver harmoniosamente, enquanto uma profissão que se coloca na relação entre a APS e a vida das pessoas: *"É engolir sapos (...) para ter uma vivência tranquila e harmoniosa tanto dentro da equipe quanto na microárea"*.

Como desejos, os profissionais expressam um valor de reconhecimento, também de forma relacional, pois é um reconhecimento que eles buscam receber da população, de outros profissionais de saúde e dos governos: *"Abraço e firme na luta irmão. Que possamos ser melhor reconhecidos o mais breve possível"*.

## 4.3 Desafios da Agentes na APS

Identificamos uma série de dificuldades que podem ser representativas da realidade de agentes de saúde na APS e na execução de campanhas de saúde. Estes desafios são apresentados a seguir.

**Desafios no Relacionamento entre Agentes e População.** Foi possível identificar desafios no estabelecimento de vínculos e relacionamentos entre agentes e população. Neste relacionamento, surgem conflitos e problemas como demandas que a população realiza para agentes, mas que não são de sua atribuição: *"diariamente recebo questionamentos de profissionais, de pessoas da população referente a direitos e deveres"*. Agentes de saúde tentam orientar e convencer a população sobre quais são suas atribuições e responsabilidades: *"(...) não é atribuição de agentes de saúde selecionar famílias para receber o benefício 'Cesta Básica'"*. Neste caso específico, houve uma demanda de algo muito importante para a população, que é um benefício socioeconômico fornecido pelo governo. Um ACS afirmou em uma publicação que os beneficiários da bolsa família devem comparecer à UBS uma vez no semestre para acompanhamento da saúde (pelo menos até 2021). Quando os beneficiários não aparecem, os ACS devem buscar essas pessoas, verificando o que ocorreu.

Como há essa associação entre ACS e uma obrigação em saúde, as pessoas podem também entender que o ACS tem atribuições em torno de dar e manter o benefício social. Do mesmo modo, como agentes de saúde são servidores públicos, essa demanda aparece como se eles soubessem e fossem responsáveis por informar e atuar em relação a esse benefício. Quando agentes de saúde dizem não serem responsáveis, há um desencontro de expectativa de famílias e a resposta de agentes de saúde, em que pode ocorrer uma quebra de vínculos. A falta de material de identificação dos agentes (como jaleco, uniforme e crachá), além da falta de criação de vínculos pode acarretar em problemas de não conseguir ser aceito no interior das casas das famílias: "É ser gente, é ser acolhedor, é ter empatia… mesmo que seja mal recebido(a) em alguns domicílios". Em alguns casos, há desafios até para entrar em espaços como condomínios: "*Os agentes estão tendo dificuldade de entrar em alguns condomínios para levar as ações preventivas de combate ao mosquito da dengue, cadastramento da população e de pessoas idosas e acamadas*".

Ainda sobre a quebra de expectativas entre agentes de saúde e a população, muitas vezes a população solicita não só orientação, mas suporte em serviços de saúde que não são atribuição dos agentes. Foi identificado, por exemplo, que a população chama o ACE para dar suporte em situações que não são de sua atribuição: *"ainda hoje somos acionados para dar suporte em situações que não são atribuições nossas, mas que envolvem animais. A população ainda tem dificuldade de entender quais são as nossas atribuições."* Por um lado, quando os agentes não podem ajudar, pode ocorrer uma não compreensão e criar um aborrecimento da população. Por outro lado, caso os agentes realizam a ajuda mesmo não sendo sua atribuição, podem ficar sem tempo para atender outras demandas que são de fato parte de sua profissão.

Neste cenário, agentes lidam com uma população que expressa descontentamento com sua atuação. Há desabafos online em que agentes dizem ouvir que "*servidor público não trabalha*", e não percebe (e aparentemente não valoriza) o trabalho de agentes. Em um caso, uma pessoa reclamou para um agente: *"E aqui [na minha cidade] têm grupo de zoonoses ou só existe no nome porque nunca vi nas ruas."* Neste caso, o agente explicou diversos desafios que impedem que a cidadã não identifique o trabalho dos agentes:

> "*Senhora (...) em que cidade e estado a sra mora? De qualquer forma eu acredito que exista sim setor de controle de zoonoses na sua cidade. O que pode ocorrer é essa equipe estar defasada por falta de contratações da gestão, o bairro da senhora nao ter agente zoneado pelo motivo acima, deles estarem trabalhando descaracterizados por não receberem fardamento em tempo habil, não existir divulgação das atividades por eles realizadas. Etc."*

O trecho explica diversos aspectos da não visibilidade de agentes. Diversas dessas explicações se colocam como desafios para a APS no contexto dos agentes: "equipe defasada por falta de contratação", o que reduz a cobertura dos agentes nos territórios; agentes "trabalhando descaracterizados por não receberem fardamento", o problema de agentes trabalharem sem colete, crachá e outros símbolos que a população possa reconhecer esses profissionais passando pelos domicílios e território; a literatura já indica que os profissionais têm a questão da identidade com seu trabalho estreitamente ligada à posse e ao uso do crachá (Evangelista et al., 2018). "não existir divulgação das atividades por eles realizadas", o que indica a relevância dessas páginas de internet mantidas pessoalmente pelos agentes de saúde em fornecer para a população um feedback das ações que são sim realizadas por esses agentes. Essas divulgações são então um importante meio da população conhecer o trabalho dos agentes e começar a participar mais das ações que são realizadas, por exemplo, dentro das UBS. Essas ações de divulgação podem também permitir conscientizar sobre a importância e trabalho dos agentes

Com essa fala, há indícios de que a totalidade dos desafios influenciam em uma invisibilidade desses profissionais para a população. Essa falta de reconhecimento da população está relacionada com uma falta de divulgação de ações expressivas dos agentes, como em uma determinada localidade houve a vacinação de 10 mil animais contra a raiva em um único ano. Ao apresentar esse número expressivo em uma publicação, o agente completou: *"Mas é uma pena que os programas de prevenção só são notados ou viram notícia no jornal quando dão errado."*

A população também sofre em relação à comunicação em saúde, onde as informações desorganizadas ou inacessíveis implicam que a população perca as ações realizadas. Em um caso, um cidadão foi ao posto de saúde em relação a campanha do novembro azul, mas já havia acabado a ação prometida pela prefeitura e ele não conseguiu realizar o exame PSA: *"Nessa publicação da prefeitura estava marcado das 9:00 as 16:00 no posto (...) multirao de exame PSA cheguei lá as 14:15 e me disseram que já havia encerrado o que será que aconteceu fiquei sem entender nada"*.

Também foram identificados desencontros de informação que pode ocorrer entre profissionais de saúde e servidores públicos, principalmente sobre temas de interesse da população, como benefícios, eventos, entre outros: *"Essa semana deram umas cestas aqui no interior e a secretaria da assistência social chegou na comunidade com listas. E as pessoas que não estavam nas listas ela dizia que nós acs que não tinha colocado o nome"*. Isso indica que até mesmo entre profissionais relacionados ao cuidado pode haver o risco de não entender bem as atribuições de agente de saúde, algo também já apontado na literatura sobre os repetidos relatos de agentes sobre problemas de desvio de função (Scherer et al., 2024).

**Realidades socioambientais específicas.** Na atuação diretamente no território em que as pessoas trabalham, estudam, se relacionam e vivem, os agentes de saúde conhecem e estão expostos à realidade desafiadora de alguns espaços socioambientais. Foi possível identificar, por exemplo, dificuldades de um ACE na atuação com populações ribeirinhas em um cenário de cheia de rio, com fotos dentro de canoas indo até domicílios realizar atendimentos: *"Vida de ACE vejam um pouquinho da dificuldade do nosso povo em relação a cheia para cuidar de seus animais"*. Em outro contexto de ACE, agora na zona rural, fotos exemplificavam lugares ermos, estradas de terra esburacadas: *"Caminhos da prevenção!! Caminhos que muitas vezes somos os primeiros ou os únicos a levarem os serviços públicos"*.

Em regiões próximas e urbanas já há desafios relacionados a condições ambientais adversas (*"a gente sai pra trabalhar com o sol tinindo na cabeça e volta correndo ensopado pela chuva (...) depois fui verificar se os papéis da sacola tinham escapado secos e colocar a mochila pra secar"*). Em lugares distantes, conseguir chegar no local onde as pessoas vivem por si só já é um desafio de agentes em regiões únicas do Brasil. A distância demanda que agentes engajem outros meios de transporte, seja individualmente (*"Vamos *de bike porque a distância é grande"*) ou em equipes, como foi possível identificar fotos de ACE utilizando motos e outros indo de Kombi para comunidades rurais.

Essas realidades socioambientais geram demandas não só para o transporte, mas para a segurança dos agentes *"para percursos longos deveria ser disponibilizado um carro de apoio, além de executar o trabalho em dupla para maior segurança do profissional"*. A presença de outro profissional pode auxiliar a mitigar casos de emergências em que os agentes estão expostos.

**Falta de equipamentos e material de trabalho adequado.** Principalmente em comentários realizados entre agentes de saúde, foi possível conhecer uma realidade diversa no Brasil em relação aos materiais de trabalho disponíveis para agentes de saúde. Neste caso, uma interação entre agentes de diferentes localidades ilustra o problema. Um agente apresenta uma foto equipado de EPIs para preparo de material larvicida. *"Dia de cortar produto é assim todo paramentado! EPIs Ok. Equipamentos Ok. Disposição Ok. Vamos lá.". Em um comentário, um agente de outra cidade demonstra* surpresa com a presença de materiais necessários para o trabalho de ACE, pois na sua realidade não tem: *"Na sua cidade tem todo esses epi's? (...)p[oi]s na minha muitas vezes falta o básico do dia a dia."* O ACE da postagem original indica que já teve no passado um histórico de um ACE que manipulou o produto sem proteção e teve reações graves de saúde: *"Esses EPIs são para o corte e fracionamento do produto, pois tivemos colegas que tiveram reação ao manuseio do produto, sendo as reações mais graves após o contato do pó do produto com olho."*

A literatura já reconhece que é um desafio ter infraestrutura e material de trabalho adequado no trabalho de agentes (Scherer et al., 2024; Evangelista et al., 2018). Foi identificado que agentes usam materiais próprios para se deslocar, como bicicleta pessoal, em um cenário em que foi fornecido bicicleta no passado, mas há muito tempo no qual a bicicleta fornecida pelo governo não permitia mais trabalhar. Para agentes de saúde, por exemplo, identificou-se a importância de equipamentos de trabalho que demonstram a formalidade profissional que ajuda a população a reconhecer esses profissionais nos territórios. A comunicação entre os agentes apresentada ilustra que a realidade de infraestrutura de trabalho não é homogênea em todo o país. Por um lado, a falta de material implica no não reconhecimento de agentes pela população (na falta de coletes e crachás), o que impede a entrada dos agentes nos domicílios. Por outro lado, a falta de material como não fornecer EPIs para ACEs pode causar impactos mais graves, colocando a saúde dos profissionais em risco.

## 5. Considerações Finais

Este relatório apresentou uma análise netnográfica sobre a atuação de agentes de saúde no contexto da APS do Brasil, a partir de análise de publicações na rede social Facebook. Ao longo da análise, constatou-se a multiplicidade de papéis dos agentes de saúde, que vão muito além do cuidado estritamente clínico e preventivo, abrangendo ações de promoção da saúde, educação em saúde, mobilização social e fortalecimento de vínculos com a comunidade. As campanhas temáticas, o uso de materiais educativos, o diálogo com outros profissionais e a participação em equipes multidisciplinares foram identificados como estratégias fundamentais para ampliar o acesso, a qualidade e a integralidade do cuidado.

No entanto, a pesquisa também evidenciou desafios que limitam e tensionam o trabalho dos ACS e ACE. Estes vão desde dificuldades estruturais, como a falta de equipamentos, EPIs, transporte e recursos materiais, até questões relacionais e de reconhecimento profissional, incluindo barreiras de acesso a domicílios, incompreensão da população sobre o papel dos agentes e a necessidade constante de negociação de atribuições. A ausência de visibilidade das ações realizadas e a carência de divulgação sistemática dos resultados obtidos também impactam a percepção social sobre o trabalho dos agentes, o que pode diminuir o engajamento comunitário e a valorização da profissão.

Do ponto de vista teórico-metodológico, o Design Socialmente Consciente orienta para a importância de entender o contexto social, o conjunto de valores e as redes de relações estabelecidas entre as partes interessadas no domínio da saúde antes da proposição de soluções tecnológicas. Nesse sentido, a netnografia cumpriu um papel exploratório essencial: sensibilizar a equipe de pesquisa quanto aos desafios, expectativas, práticas e dinâmicas socioculturais que envolvem a atuação dos agentes de saúde. Como parte de um projeto mais amplo, os achados desta investigação netnográfica não constituem conclusões definitivas, mas insumos para o aprofundamento de estudos empíricos e o desenvolvimento de tecnologias mais adequadas ao contexto brasileiro de APS. Eles poderão subsidiar a elaboração de uma solução no contexto de criação e gerenciamento de campanhas de saúde e de recebimento de feedback de usuários do SUS que privilegiem a comunicação segura, acessível e contextualizada, assim como orientar processos de co-criação com os profissionais e usuários envolvidos.

## Referências